\documentclass[cernpreprint, atlasdraft=false, texlive=2023, UKenglish, texmf, orcidlogo]{atlasdoc}
 
\usepackage{atlaspackage}
\usepackage{subfig}
\usepackage[compat=1.0.0]{tikz-feynman}
\usepackage{atlasbiblatex}
 
\usepackage{atlasphysics}[snippets=true]
\usepackage{atlasmisc}
\usepackage{MC_snippets-defs}

\graphicspath{{logos/}{figures/}}
 
\usepackage{ANA-EXOT-2020-18-PAPER-defs}

\AtlasTitle{Search for excited \ensuremath{\tau}-leptons and leptoquarks in the final state with \ensuremath{\tau}-leptons and jets in \textit{pp} collisions at $\sqrt{s} = 13~\tev$ with the ATLAS detector}

\AtlasAbstract{
A search is reported for excited \ensuremath{\tau}-leptons and leptoquarks in events with two hadronically decaying \ensuremath{\tau}-leptons and two or more jets.
The search uses proton--proton (\textit{pp}) collision data at $\sqrt{s} = 13~\tev$ recorded by the ATLAS experiment during the Run~2 of the Large Hadron Collider in 2015--2018.
The total integrated luminosity is 139~fb$^{-1}$.
The excited \ensuremath{\tau}-lepton is assumed to be produced and to decay via a four-fermion contact interaction into an ordinary \ensuremath{\tau}-lepton and a quark--antiquark pair.
The leptoquarks are assumed to be produced in pairs via the strong interaction, and each leptoquark is assumed to couple to a charm or lighter quark and a \ensuremath{\tau}-lepton.
No excess over the background prediction is observed.
Excited \ensuremath{\tau}-leptons with masses below 2.8~\tev are excluded at 95\% CL in scenarios with the contact interaction scale $\Lambda$ set to 10~\tev.
At the extreme limit of model validity where $\Lambda$ is set equal to the excited \ensuremath{\tau}-lepton mass, excited \ensuremath{\tau}-leptons with masses below 4.6~\tev are excluded.
Leptoquarks with masses below 1.3~\tev are excluded at 95\% CL if their branching ratio to a charm quark and a \ensuremath{\tau}-lepton equals 1.
The analysis does not exploit flavour-tagging in the signal region.
}

\PreprintIdNumber{CERN-EP-2023-008}

\AtlasJournal{\JHEP}
\AtlasJournalRef{JHEP 06 (2023) 199}
\AtlasDOI{DOI:10.1007/JHEP06(2023)199}

\hypersetup{pdftitle={ATLAS document},pdfauthor={The ATLAS Collaboration}}
 
\begin{document}
 
\maketitle

% The next lines are included from the .//source/intro.tex input file

\section{Introduction}
\label{sec:intro}
 
\begin{figure}
\centering
\includegraphics[width=0.45\textwidth]{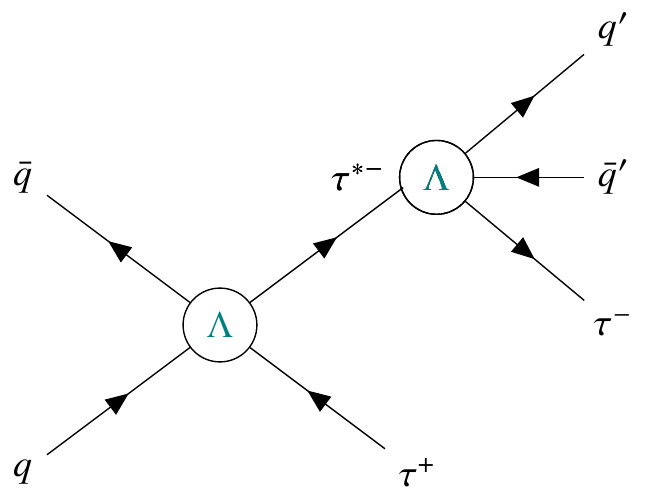}
 
\caption{Feynman diagram for \etl production and decay.
The compositeness scale below which Eq.~\ref{eq:CI} holds is denoted by   $\Lambda$.
}
\label{fig:ci}
\end{figure}

The quarks and leptons in the Standard Model (SM) could be composed of more fundamental particles.
The constituents are called preons in a model of composite quarks and leptons by Baur, Spira, and Zerwas~\cite{PhysRevD.42.815}.
The model predicts the existence of excited states towering over the known SM leptonic and quark ground states.
A transition of the excited leptons into the ordinary ones would proceed either via interaction with SM gauge bosons (Gauge Interaction, or GI) or via a new type of interaction.
In the present analysis, an effective four-fermion contact interaction (CI) is used.
After simplification~\cite{PhysRevD.42.815}, the CI interaction Lagrangian reads:
\begin{equation}
\label{eq:CI}
\mathcal{L}_\text{CI} = \frac{\left(4\pi\right)^2}{\Lambda^2} \frac{1}{2} j^\mu j_\mu
\end{equation}
where
\begin{equation}
\label{eq:current}
j_\mu = \bar{f}_L \gamma_\mu f_L + \bar{f}^\ast_L \gamma_\mu f^\ast_L + \bar{f}^\ast_L \gamma_\mu f_L + \text{h.c.}
\end{equation}
and $\Lambda$ is a compositeness scale below which Eq.~\ref{eq:CI} holds.
In Eq.~\ref{eq:current}, $f_L$ and $f^\ast_L$ stand for left-handed components of ordinary and excited fermion fields, respectively.
In \textit{pp}~interactions at the LHC, the GI plays a negligible role in the excited lepton production for $\elm>300$~\GeV and $\Lambda\lessapprox 15$~\TeV,
the range the analysis reported here focuses on.
Both the GI and CI are important in excited lepton decays~\cite{EXOT-2017-22}.
However, the CI decays dominate for values of $\elm/\Lambda$ larger than 0.1--0.3 depending on the model parameters~\cite{CMS-EXO-18-013}.
As $\elm/\Lambda$ nears one, the effective four-fermion CI description becomes inaccurate, with a severity that depends on the underlying physics~\cite{PhysRevD.42.815}.
The weaker the coupling between an excited lepton and SM gauge bosons, the lower the importance of GI decays.
In the analysis reported here, the GI couplings are assumed to be zero.
The only non-zero CI term considered is the CI between two quarks and two leptons.
The focus of the present search is on the process represented in Figure~\ref{fig:ci}.
Previous searches for excited tau leptons, \etl, were done at LEP, with the ALEPH~\cite{Barate:346643}, DELPHI~\cite{DELPHI:2004ili}, L3~\cite{Achard_2003}, and
OPAL~\cite{OPAL:2002wkp} experiments, and at the LHC, with data collected by the ATLAS experiment at 8~\TeV~\cite{EXOT-2012-20}.
The last excluded the existence of \etl with mass below 2.5~\TeV for a scenario in which the compositeness scale $\Lambda$ is equal to the \etl mass.
The study focused on \etl predicted by the same model~\cite{PhysRevD.42.815} as in the current paper, but both the GI and CI decays were considered.
However, it is possible to compare these results with the analysis reported here in the regime of $\elm/\Lambda$ values close to one or negligible \etl coupling to the SM gauge bosons.
In this regime, the CI decays dominate.
 
\begin{figure}
\centering
\includegraphics[width=0.45\textwidth]{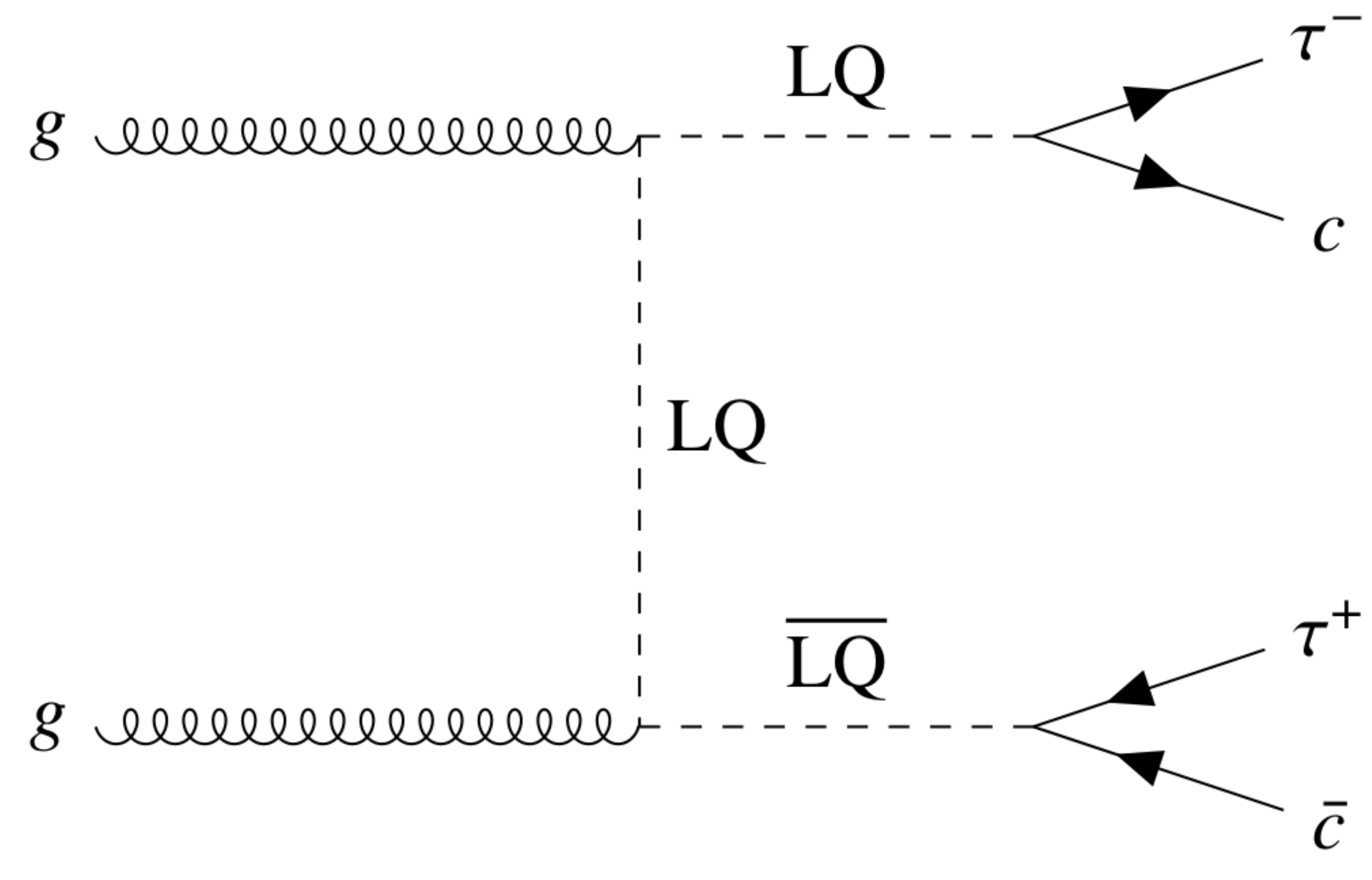}
\caption{An example Feynman diagram of the LQ pair production and decay in \textit{pp}~collisions at the LHC.}
\label{fig:LQprod}
\end{figure}

Leptoquarks (LQs) are hypothetical particles predicted by many extensions of the SM~\cite{BSM,techni2,techni3,string,comp,pati_salam_colour,georgi_glashow_unification,Diaz:2017lit}.
Each LQ simultaneously couples to a lepton and a quark, hence its name.
Scalar LQs couple to a lepton--quark pair via a Yukawa-type interaction~\cite{LQlq_coupling}, with a Yukawa coupling constant denoted by $\lambda$.
LQs can also have spin~1, but this scenario is not considered in this analysis.
LQs are colour-triplet particles, and they carry a fractional electric charge.
Due to their colour charge, they can be produced via the strong interaction in \textit{pp}~collisions at the LHC, as depicted in Figure~\ref{fig:LQprod}.
The LQ pair production is mostly insensitive to the Yukawa coupling, $\lambda$,~\cite{Diaz:2017lit} in the mass range considered.
Many models assume that LQs can couple just to one specific fermion generation.
Such an assumption is made in the minimal Buchm\"uller--R\"uckl--Wyler (BRW) model~\cite{Buchmuller:1986zs}.
The focus in this work is on the search for a LQ that couples to a \tl-lepton and a $c$-quark, as suggested in Ref.~\cite{Diaz:2017lit}.
The assumed branching ratio to this fermion pair is one.
It is the first ATLAS search for such a LQ\@.
A cross-generation LQ coupling is motivated by recent anomalies in measurements of $R(D)$ and $R(D^*)$ testing lepton flavour universality in low energy experiments~\cite{BaBar:2013mob,Belle:2015qfa,LHCb:2015gmp,Belle:2016dyj,Belle:2016ure,LHCb:2017vlu}.
If the deviations persist, they are explainable by a particular LQ type coupling to the $c$-quark--\tl-lepton pair~\cite{Freytsis:2015qca}.
Given that the analysis does not use flavour-tagging of jets in the signal region and the event kinematics are very similar if lighter
quarks are considered, the results presented here hold for scenarios in which LQs couple to a \tl-lepton and a $u$-, $d$- or $s$-quark, as well.
A similar search conducted by CMS~\cite{CMS:2018iye} using a benchmark model with a LQ coupling to a \tl-lepton and a $b$-quark resulted in a lower mass limit of 1.02~\tev.
However, similar to this study, the CMS analysis does not use jet flavour-tagging in the signal region and is also sensitive to the LQ decay study reported here.

% End of text imported from the .//source/intro.tex input file

% The next lines are included from the .//source/detector.tex input file
\newcommand{\AtlasCoordFootnote}{
ATLAS uses a right-handed coordinate system with its origin at the nominal interaction point (IP)
in the centre of the detector and the \(z\)-axis along the beam pipe.
The \(x\)-axis points from the IP to the centre of the LHC ring,
and the \(y\)-axis points upwards.
Cylindrical coordinates \((r,\phi)\) are used in the transverse plane,
\(\phi\) being the azimuthal angle around the \(z\)-axis.
The pseudorapidity is defined in terms of the polar angle \(\theta\) as \(\eta = -\ln \tan(\theta/2)\).
Angular distance is measured in units of \(\Delta R \equiv \sqrt{(\Delta\eta)^{2} + (\Delta\phi)^{2}}\).}
 
\section{ATLAS detector}
\label{sec:detector}
 
The ATLAS detector~\cite{PERF-2007-01} at the LHC is a multipurpose particle physics detector with a forward–backward symmetric cylindrical geometry and nearly $4\pi$ coverage in solid angle.\footnote{\AtlasCoordFootnote}
It consists of an inner tracking detector surrounded by a thin superconducting solenoid, electromagnetic and hadron calorimeters,
and a muon spectrometer incorporating three large superconducting air-core toroidal magnets.
 
The inner-detector system is immersed in a \qty{2}{\tesla} axial magnetic field
and provides charged-particle tracking in the range of \(|\eta| < 2.5\).
The high-granularity silicon pixel detector covers the vertex region and typically provides four measurements per track,
the first hit normally being in the insertable B-layer (IBL) installed before Run~2~\cite{ATLAS-TDR-19,PIX-2018-001}.
It is followed by the silicon microstrip tracker (SCT), which usually provides eight measurements per track.
These silicon detectors are complemented by the transition radiation tracker (TRT),
which enables radially extended track reconstruction up to \(|\eta| = 2.0\).
The TRT also provides electron identification information
based on the fraction of hits (typically 30 in total) above a higher energy-deposit threshold corresponding to transition radiation.
 
The calorimeter system covers the pseudorapidity range \(|\eta| < 4.9\).
In the region \(|\eta|< 3.2\), electromagnetic calorimetry is provided by barrel and
endcap high-granularity lead/liquid-argon (LAr) calorimeters,
with an additional thin LAr presampler covering \(|\eta| < 1.8\)
to correct for energy loss in material upstream of the calorimeters.
Hadron calorimetry is provided by the steel/scintillator-tile calorimeter,
segmented into three barrel structures in the region \(|\eta| < 1.7\), and two copper/LAr hadron endcap calorimeters.
The solid angle coverage is completed with forward copper/LAr and tungsten/LAr calorimeter modules
optimised for electromagnetic and hadronic energy measurements respectively.
 
The muon spectrometer (MS) comprises separate trigger and
high-precision tracking chambers measuring the deflection of muons in a magnetic field generated by the superconducting air-core toroidal magnets.
The field integral of the toroids ranges between \num{2.0} and \qty{6.0}{\tesla\metre}
across most of the detector.
Three layers of precision chambers, each consisting of layers of monitored drift tubes, cover the region \(|\eta| < 2.7\),
complemented by cathode-strip chambers in the forward region, where the background is highest.
The muon trigger system covers the range \(|\eta| < 2.4\) with resistive-plate chambers in the barrel, and thin-gap chambers in the endcap regions.
 
Events are selected by the first-level trigger system implemented in custom hardware,
followed by selections made by algorithms implemented in software in the high-level trigger~\cite{TRIG-2016-01}.
The first-level trigger accepts events from the \qty{40}{\MHz} bunch crossings at a rate below \qty{100}{\kHz},
which the high-level trigger reduces further to record events to disk at about \qty{1}{\kHz}.
 
An extensive software suite~\cite{ATL-SOFT-PUB-2021-001} is used in data simulation, in the reconstruction
and analysis of real and simulated data, in detector operations, and in the trigger and data acquisition
systems of the experiment.

% End of text imported from the .//source/detector.tex input file

% The next lines are included from the .//source/datamc.tex input file
\begingroup
\let\clearpage\relax
 
\section{Data and Monte Carlo samples}
\label{sec:datamc}
 
\endgroup
 
The analysis uses $\sqrt{s} = 13~\tev$ \textit{pp}~collisions data recorded by the ATLAS experiment in 2015--2018.
The integrated luminosity of the data sample is 139~\ifb.
The uncertainty in the combined 2015--2018 integrated luminosity is 1.7~\%~\cite{ATLAS-CONF-2019-021}, obtained using the LUCID-2 detector~\cite{LUCID2} for the primary luminosity measurements.
The data sample is selected by requiring good conditions for the beams and the ATLAS detector.
 
Monte Carlo (MC) simulation is used to model signal and background events.
Each generated event is processed with a detailed simulation of the ATLAS detector response to particles.
For background samples, the simulation is entirely based on \GEANT~\cite{Agostinelli:2002hh}.
The signal MC samples are processed with a fast simulation~\cite{SOFT-2010-01} which relies on a parameterisation of the
calorimeter response~\cite{ATL-PHYS-PUB-2010-013}.
 
The effect of multiple interactions in the same and neighbouring bunch
crossings (\pileup) is modelled by overlaying the simulated hard-scattering event with
inelastic \textit{pp}~events generated with \PYTHIA[8.186]~\cite{Sjostrand:2007gs}
using the \NNPDF[2.3lo] set of parton distribution functions (PDF)~\cite{Ball:2012cx} and the
A3 set of tuned parameters~\cite{ATL-PHYS-PUB-2016-017}.
The MC events are weighted to reproduce the
distribution of the average number of interactions per bunch crossing
(\(\left<\mu \right>\)) observed in the data. The \(\left<\mu \right>\)
value in data is rescaled by a factor of \(1.03\pm 0.04\) to improve
agreement between data and simulation in the visible inelastic
\textit{pp}~cross-section~\cite{STDM-2015-05}.

\subsection{Signal MC samples}
\label{sec:datamcsig}
 
Signal samples simulating the \etl production are generated with the \PYTHIA[8.243]~\cite{Sjostrand:2014zea} MC generator
with parameters set according to the A14 tune~\cite{ATL-PHYS-PUB-2014-021} and using the \NNPDF[2.3lo]
set of PDFs~\cite{Ball:2012cx}.
The \EVTGEN~\cite{Lange:2001uf} generator interfaced with Pythia is used to simulate decays of unstable particles and to model spin correlations and polarisation of particles in decays of hadrons.
The \etl are produced via CI in \qqbar scattering.
The same interaction governs their decays into a \tl-lepton and a \qqbar pair.
In the decays, all quark flavours are considered when kinematically allowed.
The MC generator is configured with the compositeness scale $\Lambda=10$~\TeV, but the result is also interpreted for other values.
The signal samples are produced with steps of 100~\GeV for masses between 400~\GeV and 1~\TeV and steps of 250~\GeV between 1~\tev and 10~\tev.
Uncertainties in the \etl MC predictions are taken from PDF variations.
The NNPDF~\cite{Ball:2012cx} uncertainty is determined using  100 replicas~\cite{Bothmann:2016nao}
provided by the LHAPDF6 tool~\cite{Buckley:2014ana}.
A second PDF uncertainty component is derived from the differences between predictions obtained with \NNPDF[2.3lo]~\cite{Ball:2012cx}, \MMHT[lo68cl]~\cite{Harland-Lang:2014zoa} and \CT[14lo]~\cite{Dulat:2015mca}.

Simulated events with pair-produced scalar LQs are generated at next-to-leading-order (NLO) in quantum chromodynamics (QCD) with \MGNLO[2.6.0]~\cite{Alwall:2014hca},
using the method described in Ref.~\cite{Mandal:2015lca}, in which fixed-order NLO QCD calculations~\cite{Kramer:2004df,Kramer:1997hh} are interfaced to \PYTHIA[8.230]~\cite{Sjostrand:2014zea}
for the parton shower (PS) and hadronisation.
Parton luminosities are provided by the five-flavour scheme \NNPDF[3.0nlo] set of PDFs~\cite{Ball:2014uwa} with $\alpha_\text{S}=0.118$
and the underlying event (UE) is modelled with the A14 tune~\cite{ATL-PHYS-PUB-2014-021}.
MadSpin~\cite{Artoisenet:2012st} is used for the decay of the scalar LQ\@.
The coupling parameter $\lambda$ is set to~0.3, resulting in a LQ width of about~0.2\% of its mass~\cite{Buchmuller:1986zs,LQlq_coupling}.
The charge of the LQ is set to $-1/3e$, and the $\overline{\mathrm{LQ}}$ has the opposite charge.
The LQ is assumed to couple to just one lepton--quark pair, namely $\tl-c$.
The kinematics stays the same if an $s$-quark is used instead of a $c$-quark.
Therefore, any results  are applicable to LQs coupling to a light quark and a \tl-lepton as well.
The signal samples are produced for masses between 500~\GeV and 1.7~\TeV in steps of 100~\GeV.
Contributions from higher-order corrections in perturbation theory are estimated by varying the renormalisation, \mur, and the factorisation, \muf, scales~\cite{Bothmann:2016nao}.
The NNPDF~\cite{Ball:2014uwa} uncertainty is determined using  100 replicas~\cite{Bothmann:2016nao}.
A second PDF uncertainty component is derived from the differences between predictions obtained with \NNPDF[3.0nlo]~\cite{Ball:2014uwa}, \MMHT[nlo68cl] \cite{Harland-Lang:2014zoa} and \CT[14] \cite{Dulat:2015mca}.
The uncertainty in $\alphas$ is obtained by varying  $\alphas$ by $\pm 0.001$ from the nominal value ($\alphas = 0.118$).
The initial state radiation (ISR) uncertainty is estimated by varying the Var3c parameter of the A14 tune~\cite{ATL-PHYS-PUB-2017-007}.
Total cross-sections of the LQ-pair production are obtained from calculating the pair production of scalar-coloured particles~\cite{Beenakker:2016lwe,Beenakker:1997ut,Beenakker:2010nq,Beenakker:2016gmf}.
Production of the supersymmetric partners of the top quark is used in these calculations; however, since they have the same production modes as scalar LQs and their pair-production cross-section depends only on their mass, their cross-sections apply also to the model used in this analysis.
The cross-sections are computed at approximate next-to-next-to-leading-order (NNLO) in QCD with resummation of next-to-next-to-leading order logarithmic (NNLL) soft gluon terms.
The cross-sections do not include lepton $t$-channel contributions, which are neglected in Ref.~\cite{Mandal:2015lca} and may lead to corrections at the per cent level~\cite{Borschensky:2020hot}.

\subsection{Background MC samples}
\label{sec:datamcbkg}

The production of a vector boson (\(W\), \(Z\)) and jets (\(V+\)jets) is simulated with the
\SHERPA[2.2.1]~\cite{Bothmann:2019yzt}
generator using NLO matrix elements (ME) for up to two partons, and leading-order (LO) matrix elements
for up to four partons calculated with the Comix~\cite{Gleisberg:2008fv}
and \OPENLOOPS~\cite{Buccioni:2019sur,Cascioli:2011va,Denner:2016kdg} libraries. They
are matched with the \SHERPA parton shower~\cite{Schumann:2007mg} using the \MEPSatNLO
prescription~\cite{Hoeche:2011fd,Hoeche:2012yf,Catani:2001cc,Hoeche:2009rj}
using the set of tuned parameters developed by the \SHERPA authors.
The \NNPDF[3.0nnlo] set of PDFs~\cite{Ball:2014uwa} is used and the samples
are normalised to an NNLO
prediction~\cite{Anastasiou:2003ds}.
Uncertainties in the MC prediction due to higher-order corrections are estimated by varying \mur and \muf~\cite{Bothmann:2016nao}.
Uncertainties due to \alphas and the PDF choice are estimated similarly to those for the LQ MC\@.
The uncertainty due to the jet-parton matching scheme (CKKW~\cite{Stefano_Catani_2001}) is estimated by varying the nominal value of the corresponding scale of 20~\gev to 15~and 30~\gev.
The uncertainty related to the resummation scale choice is estimated by varying the nominal value by factors of 0.5 and 2~\cite{Bothmann:2019yzt}.

The production of \ttbar and single top quark ($tW$, $tb$, $tq$) events is modelled using the
\POWHEGBOX[v2]~\cite{Frixione:2007nw,Nason:2004rx,Frixione:2007vw,Alioli:2010xd}
generator at NLO with the \NNPDF[3.0nlo]~\cite{Ball:2014uwa} PDF set.
In the \ttbar sample, only the QCD production is considered.
The \hdamp parameter\footnote{The
\hdamp parameter is a resummation damping factor and one of the
parameters that controls the matching of \POWHEG matrix elements to
the parton shower and thus effectively regulates the
high-\pT radiation against which the \ttbar system recoils.} for the \ttbar sample is set
to 1.5\,\mtop~\cite{ATL-PHYS-PUB-2016-020}.  The events are interfaced
to \PYTHIA[8.230]~\cite{Sjostrand:2014zea} to model the parton shower,
hadronisation, and underlying event, with parameters set according
to the A14 tune~\cite{ATL-PHYS-PUB-2014-021} and using the \NNPDF[2.3lo]
set of PDFs~\cite{Ball:2012cx}. The decays of bottom and charm hadrons
are performed by \EVTGEN[1.6.0]~\cite{Lange:2001uf}.
In the \tw sample, the diagram removal scheme~\cite{Frixione:2008yi} is used to
remove interference and overlap with \ttbar production.
The related uncertainty is estimated by comparison with an alternative sample
generated using the diagram subtraction scheme~\cite{Frixione:2008yi,ATL-PHYS-PUB-2016-020}.
The uncertainty due to ISR is estimated by simultaneously varying \mur, \muf and the Var3c parameter of the A14 tune~\cite{ATL-PHYS-PUB-2017-007}.
The PDF uncertainty is derived from 30 PDF variations according to the \PDFforLHC~\cite{Butterworth:2015oua} recommendations.
The impact of using a different parton shower and hadronisation model is evaluated
by comparing the nominal \ttbar or single-top quark samples with other event samples produced with the
\POWHEGBOX[v2]~\cite{Frixione:2007nw,Nason:2004rx,Frixione:2007vw,Alioli:2010xd}
generator using the \NNPDF[3.0nlo]~\cite{Ball:2014uwa} PDF set.
Events in the latter sample are interfaced with \HERWIG[7.04]~\cite{Bahr:2008pv,Bellm:2015jjp},
using the H7UE set of tuned parameters~\cite{Bellm:2015jjp} and the
\MMHT[lo] PDF set~\cite{Harland-Lang:2014zoa}.
To assess the uncertainty in the matching of NLO matrix elements to the
parton shower, the nominal \ttbar sample is compared with a sample
generated with the \MGNLO[2.6.0] generator at NLO in QCD using
the \NNPDF[3.0nlo] PDF set.
For single-top samples, the \MGNLO[2.6.2]~\cite{Alwall:2014hca} generator with the
\NNPDF[2.3nlo]~\cite{Ball:2014uwa} PDF set was used instead.
The events were interfaced with \PYTHIA[8.230]~\cite{Sjostrand:2014zea}, using the A14
set of tuned parameters~\cite{ATL-PHYS-PUB-2014-021} and the \NNPDF[2.3lo] PDF set~\cite{Ball:2012cx}.
As previous analyses~\cite{TOPQ-2015-06,ATLAS:2022xfj} demonstrated, the modelling of the background produced by top quarks can be improved by correcting all \ttbar samples to match their top quark \pt, \ttbar mass and \ttbar \pt distributions to those predicted at NNLO  in QCD and NLO  in EW\@.
Top quark pair differential calculations from Ref.~\cite{Czakon:2017wor} are used for this correction.
The reweighting is done sequentially in the three variables.
Uncertainties in the reweighting are due to PDFs, including the photon PDF, QCD scale variations, modelling of more than two QCD emissions, and the order in which the three spectra are reweighted.
The full NLO EW correction is taken as uncertainty as well.

Samples of diboson (\VV) final states are simulated with the
\SHERPA[2.2.1] or 2.2.2~\cite{Bothmann:2019yzt} generator depending on the process,
including off-shell effects and Higgs boson contributions, where appropriate.
Fully leptonic final states and semileptonic final states, where one boson
decays leptonically and the other hadronically, are generated using
matrix elements at NLO  in QCD for up to one additional parton
and at LO  for up to three additional parton
emissions. Samples for the loop-induced processes \(gg \to VV\) are
generated using LO matrix elements for up to one
additional parton emission for both the fully leptonic and semileptonic final states.
The matrix element calculations are matched
and merged with the \SHERPA parton shower based on Catani--Seymour
dipole factorisation~\cite{Gleisberg:2008fv,Schumann:2007mg} using the MEPS@NLO
prescription~\cite{Hoeche:2011fd,Hoeche:2012yf,Catani:2001cc,Hoeche:2009rj}.
The virtual QCD corrections are provided by the
\OPENLOOPS library~\cite{Buccioni:2019sur,Cascioli:2011va,Denner:2016kdg}. The
\NNPDF[3.0nnlo] set of PDFs is used~\cite{Ball:2014uwa}, along with a dedicated set of tuned parton-shower parameters developed by the
\SHERPA authors.
The only uncertainty considered is due to missing higher-order corrections, which are estimated similarly to those for the $V+$jets.
 
Table~\ref{tab:mcsamples} shows a summary of the MC samples used to simulate the main signal and background processes.
 
\begin{table}[t]
\centering
\caption{Overview of the main MC samples used for the signal and background simulation.
\AtlasMC[2.6.0]{MG5\_aMC} is an abbreviation for \MGNLO[2.6.0].
}
\begin{tabular}{lcccc}
\hline
Sample & \multicolumn{2}{c}{Generator} & PDF & Tune \\
& ME & PS & & \\
\hline
Signal \etl & \multicolumn{2}{c}{\PYTHIA[8.243]} & \NNPDF[2.3lo] & A14 \\
Signal LQ & \AtlasMC[2.6.0]{MG5\_aMC} & \PYTHIA[8.230] & \NNPDF[3.0nlo] & A14 \\
$V+$jets & \multicolumn{2}{c}{\SHERPA[2.2.1]} & \NNPDF[3.0nnlo] & \SHERPA \\
\ttbar & \POWHEGBOX[v2] & \PYTHIA[8.230] & \NNPDF[3.0nlo] & A14 \\
Single-top & \POWHEGBOX[v2] & \PYTHIA[8.230] & \NNPDF[3.0nlo] & A14 \\
Diboson & \multicolumn{2}{c}{\SHERPA[2.2.1] or 2.2.2} & \NNPDF[3.0nnlo] & \SHERPA \\
\hline
\end{tabular}
\label{tab:mcsamples}
\end{table}

% End of text imported from the .//source/datamc.tex input file

% The next lines are included from the .//source/eventreco.tex input file
 
\section{Event reconstruction}
\label{sec:eventreco}
 
In the data sample used for the present analysis, the number of \textit{pp}~collisions occurring simultaneously in a bunch crossing varies from about 20 to 70.
The locations of individual \textit{pp}~collisions are called vertices, and each of them is reconstructed from at least two tracks with $\pt > 500~\mev$.
Additional requirements on the tracks guarantee that they originate in a region where the beams overlap in the transverse plane.
The primary vertex is defined as the vertex with the largest sum of squared \pt of its matched tracks~\cite{ATL-PHYS-PUB-2015-026}.
 
Each electron candidate consists of a track matched to a cluster of energy deposited in the electromagnetic calorimeter~\cite{PERF-2017-01}.
Selection criteria for the track impact parameters are imposed to guarantee that the track originates at the primary vertex.
The transverse, $d_0$, and longitudinal, $z_0$, impact parameters are required to satisfy $d_0 / \sigma\left(d_0\right) < 5$ and $|z_0 \sin\theta| < 5~\text{mm}$ where $\sigma\left(d_0\right)$ is the uncertainty in $d_0$ and $\theta$ is the track's polar angle.
The pseudorapidity of the calorimeter energy cluster must satisfy $|\eta_\text{cluster}| < 1.37$ or $1.52 < |\eta_\text{cluster}| < 2.47$.
The electron candidates are required to have $\pt > 15~\gev$ and to be identified as electrons using the \textit{Loose} selection criterion described in Ref.~\cite{EGAM-2018-01}.
A multivariate algorithm further suppresses non-prompt electrons from $b$-hadron decays.
It is an improved version of the isolation technique described in Ref.~\cite{HIGG-2017-02}.
 
Each muon candidate consists of a track reconstructed in the inner detector matched to a track found in the muon spectrometer.
Information about the two tracks is combined to get a more precise measurement of the muon momentum~\cite{PERF-2015-10}.
The resulting track is then required to satisfy the same criteria for the impact parameters as electron tracks.
Finally, muon candidates are selected if they have $\pt > 7~\gev$, $|\eta|<2.5$ and  satisfy the \textit{High-\pt} identification (ID) described in Ref.~\cite{PERF-2015-10}.
Muons are also required to satisfy similar lepton isolation criteria as electrons.
 
Hadronically decaying \tl-lepton candidates, \tauhad, are reconstructed from energy clusters in the calorimeters
and matched inner detector tracks~\cite{ATLAS-CONF-2017-029,ATL-PHYS-PUB-2015-045,PERF-2013-06}.
The transverse momentum of \tauhad cannot be fully reconstructed because of an undetected neutrino.
The energy scale of the visible decay products of the \tauhad is measured using $Z \rightarrow \mu \tauhad 3\nu$ events.
Five categories (decay modes) of hadronic decays are reconstructed: decays with one matched track and either zero, one or more neutral particles, and decays with three matched tracks and either zero or more neutral particles.
Only $\tauhad$s with reconstructed electric charge $|q| = 1$ are selected.
Each \tauhad is required to have $\pt > 20~\gev$, $|\eta|<1.37$ or $1.52<|\eta|<2.5$ and to have a score greater than 0.01 from an ID algorithm based on Recurrent Neural Networks (RNN)~\cite{ATL-PHYS-PUB-2019-033}.
In the MC samples, each \tauhad candidate is required to geometrically match the generator-level \tl, electron, or muon particle.
The requirement aims to remove jets (from the hadronisation of quarks or gluons) misidentified as \tauhad from the MC samples.
Background from jets misidentified as \tauhad is estimated
by using a data-driven technique (see Section~\ref{sec:bkgfake}).
However, background from light leptons misidentified as $\tauhad$ is estimated by using the MC.
 
Jets are reconstructed from constituents built according to the particle flow algorithm that exploits both the tracks
and calorimeter energy clusters~\cite{PERF-2015-09,JETM-2018-05}.
The particle flow objects are passed to the \antikt algorithm~\cite{Cacciari:2008gp,Fastjet} with a radius parameter of \(R=0.4\) to form jets with a four-momentum recombination scheme.
Jet energy is calibrated to the hadronic scale with the effect of \pileup removed~\cite{JETM-2018-05}.
In situ jet calibration consists of measurements with $Z+$jets, $\gamma+$jets and multijet events, and it is also used to derive the jet energy scale (JES) uncertainty.
Jets are required to have $\pt > 20~\gev$ and $|\eta| < 4.5$.
 
As the data are reconstructed by independent algorithms for each type of object, it is common for the same detector energy depositions to be reconstructed as several different types of objects.
For example, a \tauhad is almost always reconstructed as a jet.
It is necessary to remove this ambiguity and keep each object only once.
The procedure for removing the redundant reconstructed objects is described in Ref.~\cite{ATLAS:2022yrq}.
 
A multivariate $b$-tagging algorithm is used to tag jets as originating from a $b$-quark~\cite{ATLAS:2022qxm,FTAG-2018-01,ATL-PHYS-PUB-2017-013}.
It exploits information about the impact parameters of displaced tracks and properties of vertices in the jets.
The algorithm tags jets with $\pt > 20~\gev$ and $|\eta| < 2.5$.
A fixed 85\% efficiency working point is used to select top-quark-enriched events to estimate this background.
 
Another multivariate algorithm, the Jet Vertex Tagger (JVT)~\cite{PERF-2014-03}, is used to veto jets with $\pt<60~\gev$ and $|\eta|<2.5$ if their constituent tracks are not consistent with the primary event vertex.
Similarly, a forward Jet Vertex Tagging (fJVT)~\cite{PERF-2016-06-witherratum} algorithm vetoes jets with $\pt<60~\gev$ and $|\eta|>2.5$.
If a jet fails either of the two algorithms, then it is rejected.
 
The missing transverse momentum \ptmis is defined as the negative vectorial sum of transverse momenta of all objects in the event.
Soft particle tracks not matched to any object are taken into account via a separate 'soft term'~\cite{ATL-PHYS-PUB-2015-027}.
The magnitude of the \ptmis is referred to as \etmis.

% End of text imported from the .//source/eventreco.tex input file

% The next lines are included from the .//source/strategy.tex input file
 
\section{Search strategy}
\label{sec:strategy}
 
A search is made for two hypothetical particles: a \etl, produced and decayed in the process shown in Figure~\ref{fig:ci}, and a LQ, coupling to a \tl-lepton and a $c$-quark.
The LQ is produced in a particle-antiparticle pair (see Figure~\ref{fig:LQprod}).
Both models predict an excess of events with two \tl-leptons and two jets over the SM expectation.
In the present analysis, no jet flavour-tagging is used in the signal region, and only hadronic \tl-lepton decays are considered.
{\etl}s and LQs with masses above 300~\GeV and 500~\GeV, respectively, are considered in this search; therefore, the number of events with high-momentum jets and $\tauhad$s is enhanced in both signal types.
The analysis signal region (SR) is designed to select these events.
A particularly sensitive variable to the signal is the scalar sum of transverse momenta of the two leading-\pt jets and the two $\tauhad$s, referred to as \st.
 
The main backgrounds are events containing a $Z$~boson that decays into \tl-leptons, \ttbar and single top quark events (together referred to as Top), and Fake \tauhad events.
Fake \tauhad events contain at least one jet misidentified as \tauhad, and their yields are estimated from data.
The shapes of the kinematic distributions of the \ztt and Top backgrounds are estimated from MC, while their total yield is determined from data.
Dedicated \ztt and Top control regions (CRs) are used to improve the MC predictions.
In addition, small backgrounds such as \zll (with $\ell$ being an electron or a muon), diboson, and $W$ production are also considered.
These backgrounds contain events where one or two light leptons (mainly electrons) are misidentified as \tauhad.
These events are treated in the same way as backgrounds with real \tl-leptons.
 
The hypothesised existence of {\etl}s and LQs is probed in a statistical test based on a profile likelihood ratio test statistic.
Background and signal templates are binned in \st in the SR and both the CRs.
Systematic uncertainties are implemented in the likelihood function as parameters with a constraint term (nuisance parameters, or NPs).
The NPs control the shape and normalisation of the MC templates used in the binned maximum-likelihood fit.
For each systematic uncertainty, an NP is defined so that a value of $0$ corresponds to the nominal MC prediction, the values of $\pm1$ correspond to the $\pm1\,\sigma$ systematic variation of the MC template (constructed according to the methodology prescribed for the given systematic uncertainty), and for any value in between the shape is an interpolation between the nominal and varied template.
The uncertainty due to the limited sample size of the MC templates is parameterised by NPs with Poisson constraints assigned to each \st bin of the signal and control regions.
 
In addition, the fit model has two normalisation factors (NPs without the constraint terms) which control the normalisations of the Top and \(Z\) backgrounds, where the \(Z\) background consists of all \(Z\) decays into charged leptons.
The parameter of interest of the analysis is the signal strength $\mu_\text{sig}$, the normalisation factor of the signal template.
Its value and uncertainty are derived in a simultaneous maximum-likelihood fit of the background and signal templates to data in the SR and the two CRs.
 
For background processes assigned a normalisation factor, \(Z\) and Top production, each MC variation is normalised so that its predicted total event yield in the CRs and the SR is the same as the nominal prediction.
This normalisation is used because any change in the total yield can always be absorbed into the change of the normalisation factor and has no impact on the parameter of interest.
For signal theory uncertainties, the treatment is the same, but the variation of the total yield of signal events is then displayed as an uncertainty band on the predicted cross-section in the limit plot (c.f. Section~\ref{sec:result}).
The effects of experimental uncertainties (c.f. Section~\ref{sec:bkgunc}) on the total signal event yield are kept at full magnitude.
 
The background normalisation factors are mostly constrained due to the CRs, which are designed to have a larger sample size
and higher purity of events produced by a given process -- \ztt or Top production -- than the SR.
Therefore, the fit with and without the SR yields similar background predictions.
The background modelling of $Z$, Top, and Fake \tauhad processes is validated using dedicated validation regions (VRs) for each component.
 
The likelihood function is built and handled with the HistFitter tool~\cite{Baak:2014wma} based on the HistFactory~\cite{Cranmer:2012sba}, RooStats~\cite{roostats}, RooFit~\cite{roofit} and MINUIT2~\cite{James:1994vla} frameworks.
A $p$-value is calculated using asymptotic formulae for the profile likelihood ratio test statistic distribution~\cite{Cowan:2010js}.
The CLs method~\cite{Junk:1999kv,Read:2002hq} is used to derive all limits.

\subsection{Signal Region}
\label{sec:sr}
 
The analysis SR consists of events with exactly two reconstructed $\tauhad$s and at least two reconstructed jets, as defined in Section~\ref{sec:eventreco}.
Events with an electron or a muon are vetoed.
The events are triggered with a di-\tauhad trigger, and the two reconstructed \tauhad objects must be matched to the trigger objects.
The online (offline) \pt thresholds for the two $\tauhad$s are 35 and 25~\GeV (40 and 30~\GeV).
The leading and subleading \tauhad in \pt are further required to satisfy the \textit{Medium} and \textit{Loose} RNN ID selection criteria~\cite{ATL-PHYS-PUB-2019-033}, respectively.
The efficiency of the \textit{Medium} (\textit{Loose}) requirement is 75\% (85\%) for {\tauhad}s with one track and 60\% (75\%) for {\tauhad}s with three tracks.
The two $\tauhad$s must be geometrically distant from each other, with a criterion $\Delta R > 0.8$, and have opposite signs of the reconstructed electric charge.
The collinear di-\tauhad mass~\cite{ELLIS1988221,ATLAS:2022yrq}, \mtt, is required to be larger than 110~\GeV to suppress the \ztt background.
 
The collinear approximation assumes that neutrinos travel in the same direction as the
visible \tauhad decay products and are the only source of \ptmis.
With this assumption, in the transverse plane, one can decompose the \ptmis vector into two signed components
along the directions of the leading and subleading \pt of the \tauhad visible decay products.
This provides estimates of the full \tl transverse momenta, $p_{\mathrm{T},0}^{\mathrm{coll}}$ and $p_{\mathrm{T},1}^{\mathrm{coll}}$, and the di-\tauhad invariant mass \mtt.
For each \tl decay, a visible momentum fraction, $x$, can be calculated from the reconstructed visible transverse momentum, $p_{\mathrm{T}}^{\mathrm{vis}}$, and the estimated full transverse momentum $p_{\mathrm{T}}^{\mathrm{coll}}$, as a ratio $x = p_{\mathrm{T}}^{\mathrm{vis}} / p_{\mathrm{T}}^{\mathrm{coll}}$.
If an event contains two genuine \tl-lepton decays, these fractions are typically positive for both the leading and subleading \tl-lepton decays.
However, events with fake $\tauhad$s have \ptmis pointing in a random direction, and the fractions can become negative.
To suppress events with fake $\tauhad$s, the requirements $x_0 > 0.1$ and $x_1 > 0.05$ are used for the leading and subleading \tauhad, respectively.
A selection on the scalar sum of the two reconstructed \tauhad transverse momenta, \lt, is applied: $\lt > 140~\GeV$.
Besides the requirements on \tauhad kinematic variables, the two leading jets in \pt in the event must have $|\eta| < 2.4$.
The leading and subleading jet \pt have to satisfy the conditions $\pt > 70~\GeV$ and 60~\GeV, respectively.
 
In the mass ranges studied, the acceptance-times-efficiency of the SR selection ranges between 3.0 and 7.75\% for \etl and 7.5 and 9.0\% for LQ\@.
The maximum is reached for a \etl mass of 2~\TeV and a LQ mass of 1~\TeV.
 
\subsection{\ztt Control and Validation Regions}
\label{sec:zttcrvr}
 
The \ztt CR is defined by selection criteria similar to the SR\@.
Two conditions are different to enrich the region in \ztt events, suppress signal and leave space for a statistically independent \ztt VR\@.
First, \mtt is required to satisfy $\mtt \in (70, 110)~\GeV$ to define a region around the $Z$ boson mass peak.
Second, \lt must lie in an interval from 100~\GeV to 140~\GeV.
74\% of the events in the \ztt CR originate from $Z$ boson production, and this fraction increases to 90\% in the last \st bin.
The fraction of signal events in the region is below 1\% for all signal hypotheses described in Sec.~\ref{sec:datamcsig}.
The \ztt VR is defined by $\lt > 140~\GeV$, leaving all the other selection criteria identical to the \ztt CR.
 
\subsection{Top Control and Validation Regions}
\label{sec:topcrvr}
 
Defining a Top CR consisting of di-\tauhad events is difficult.
Either there is too much signal or the Top background does not reach high enough values of \st, needed
for reasonable control and validation of the background in a phase space close to the SR\@.
Therefore, the analysis uses a region with exactly one \tauhad.
The events are selected with a single-\tauhad trigger.
The \pt thresholds for the lowest unprescaled single-\tauhad trigger have changed throughout Run~2.
In 2015 and the first part of the 2016 data, the threshold is 80~\GeV, corresponding to an offline \pt threshold of 100~\GeV.
In the second part of the 2016 data, the online (offline) threshold is 125 (140)~\GeV.
In the last part of the 2016 and 2017--2018 data, the online	(offline) threshold is 160 (180)~\GeV.
The Top CR definition also has an upper threshold on the reconstructed \tauhad \pt of 450~\GeV in order to remove events containing high-mass signal particles.
To enrich the region in Top events, at least four jets (as defined in Section~\ref{sec:eventreco}) are required in each event, out of which two must be $b$-tagged.
Suppression of Fake \tauhad events is achieved by requiring $\etmis > 150~\GeV$.
75\% of the events in the Top CR originate from Top production, and the fraction decreases to 57\% in the last \st bin.
The fraction of signal events in the region is 1\% or lower for all \etl signal hypotheses described in Sec.~\ref{sec:datamcsig}.
The highest fraction of LQ-pair events in the Top CR is 16\%, corresponding to the LQ mass of 500~\GeV.
The fraction decreases rapidly with increasing LQ mass.
The same selection criteria define the Top VR, with only the \etmis requirement being different: $\etmis \in (120, 150)~\GeV$.
 
\subsection{Fake Validation Regions}
\label{sec:fakevr}

The Fake background is validated in regions similar to the analysis regions, but with some selection criteria changed to enrich them in jets misidentified as a $\tauhad$.
A VR for Fake background in the SR is defined by changing the opposite-sign (OS) requirement to same-sign (SS) and by introducing an upper \mtt threshold of 600~\GeV to reduce signal leaking in this VR\@.
This VR is referred to as SS Fake VR\@.
To validate the Fake background in the Top CR, a Single-\tauhad Fake VR is defined by modifying the missing transverse momentum criterion to $\etmis \in (30, 50)~\GeV$.
There is no need for a Fake CR because the Fake background is estimated with a data-driven method.
More details of how the Fake background is estimated are given in Section~\ref{sec:bkgfake}.

% End of text imported from the .//source/strategy.tex input file

% The next lines are included from the .//source/bkg.tex input file
 
\section{Background and signal estimate}
\label{sec:bkg}
 
\subsection{Uncertainties in MC predictions}
\label{sec:bkgunc}
 
The three main sources of systematic uncertainties considered are theory predictions, the experimental setup, and MC statistical uncertainty.
Sources of theoretical uncertainty and their estimates are described in Section~\ref{sec:datamc} for each MC sample used in the analysis.
 
Most experimental systematic uncertainties arise from differences between data and MC\@.
With the use of supporting measurements, the MC simulation is corrected to resemble data as closely as possible (see Section~\ref{sec:eventreco}).
Systematic uncertainties from the measurements of these corrections cover the residual differences between data and MC\@.
Differences between the \tl-lepton energy scale (TES) in data and MC are estimated by using $Z \rightarrow \tau_{\mu} \tauhad$ events, measurements of the calorimeter response to single particles and comparisons between simulations using different detector geometries or \GEANT physics lists~\cite{ATLAS-CONF-2017-029}.
JES uncertainties are estimated by using $Z+$jets, $\gamma+$jets and multijet events and additional components are due to the JES extrapolation to high jet transverse momenta, pile-up JES corrections and quark-flavour effects.
Jet energy resolution uncertainties are determined using dijet events and a noise term measurement with random cones~\cite{JETM-2018-05}.
 
Between MC and data, there are also different efficiencies from the \tauhad trigger, reconstruction and ID algorithms.
The differences are corrected with event weights (scale factors, or SFs), which are functions of the number of $\tauhad$s in the
event, their \pt, $\eta$ and the number of matched tracks.
The SFs and their uncertainties are estimated by using events with a $Z$~boson or a top-quark that decays into a \tauhad~\cite{ATLAS-CONF-2017-029}.
Furthermore, SFs are derived to compensate for differences between data and MC for both the  efficiency and inefficiency
of $b$-tagging, which are estimated together with their uncertainties using \ttbar~\cite{FTAG-2018-01} events.
SFs for jet vertex tagging efficiencies are measured using $\zmm+$jets~\cite{PERF-2014-03,PERF-2016-06-witherratum} events.
 
Each experimental uncertainty NP is assumed to be 100\% correlated between the analysis bins and all background and signal processes.
The highest ranked NPs are due to the matching of NLO matrix elements to the parton shower in \tw MC predictions, and the uncertainty in the interference of the \ttbar and \tw processes.
Another highly ranked NP is due to the TES and it comes from measurements of the calorimeter response to single particles.
No NP is strongly pulled or constrained in the fit.
 
\subsection{\ztt background estimate}
\label{sec:bkgztt}
 
The \(Z\) template is built from simulated \ztt and \zll events ($\ell = e$ or $\mu$).
It is assigned one common unconstrained normalisation factor.
In the SR, 93\% of the \(Z\) background consists of \ztt events; the rest are \zll events.
The \zll contribution drops to about 4\% in the last \st bin.
Figure~\ref{fig:bkgztt} shows the post-fit \st spectra and their total uncertainties compared to data in the \ztt CR and VR.
The fit with background-only templates is performed using the SR, \ztt and Top CRs.
A good agreement between the total background prediction and data validates the modelling of the \ztt background shape.
The fitted normalisation factor is compatible with 1.
 
\begin{figure}
\centering
\subfloat[\ztt CR]{
\includegraphics[width=0.45\textwidth]{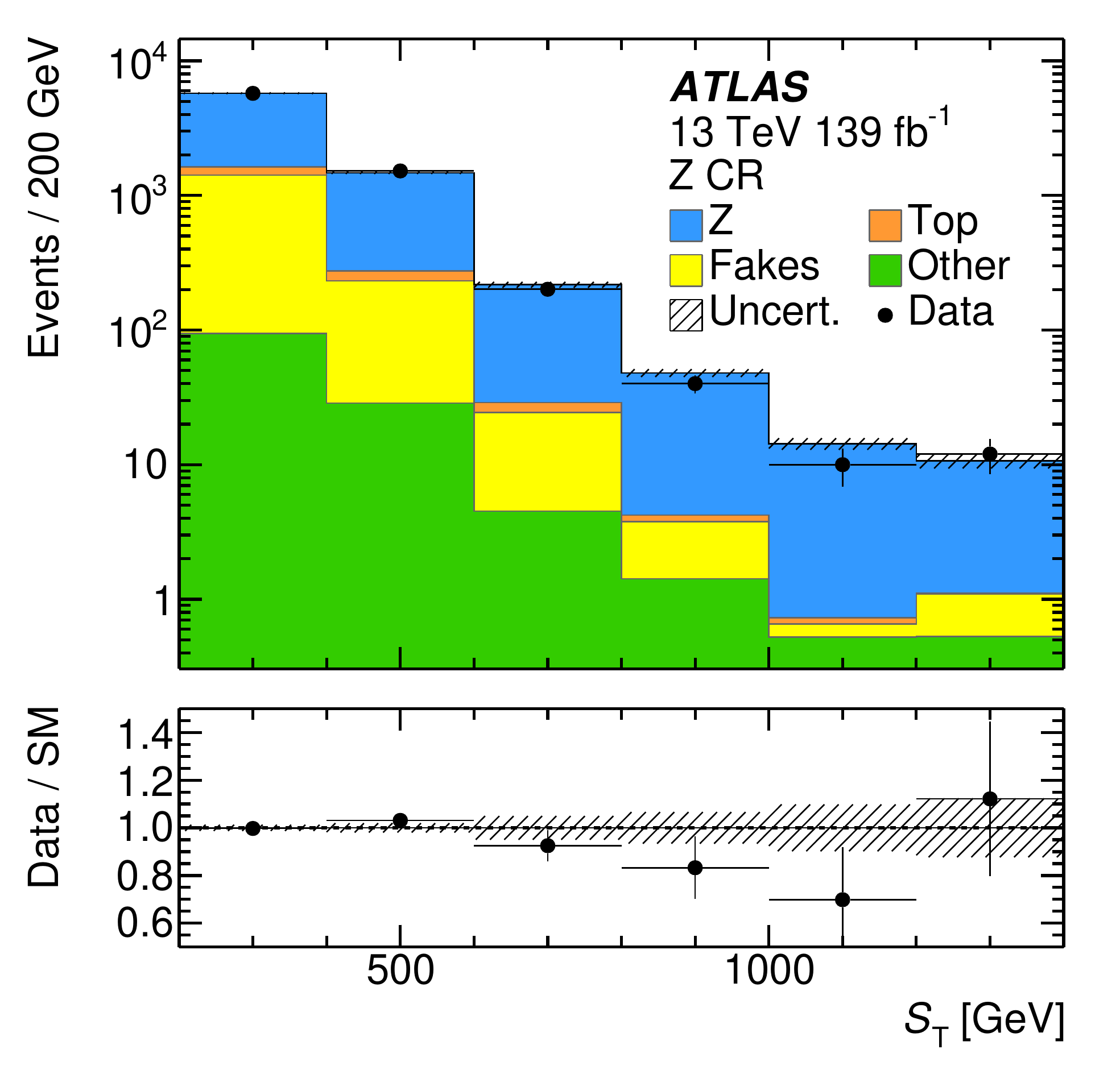}
}\quad
\subfloat[\ztt VR]{
\includegraphics[width=0.45\textwidth]{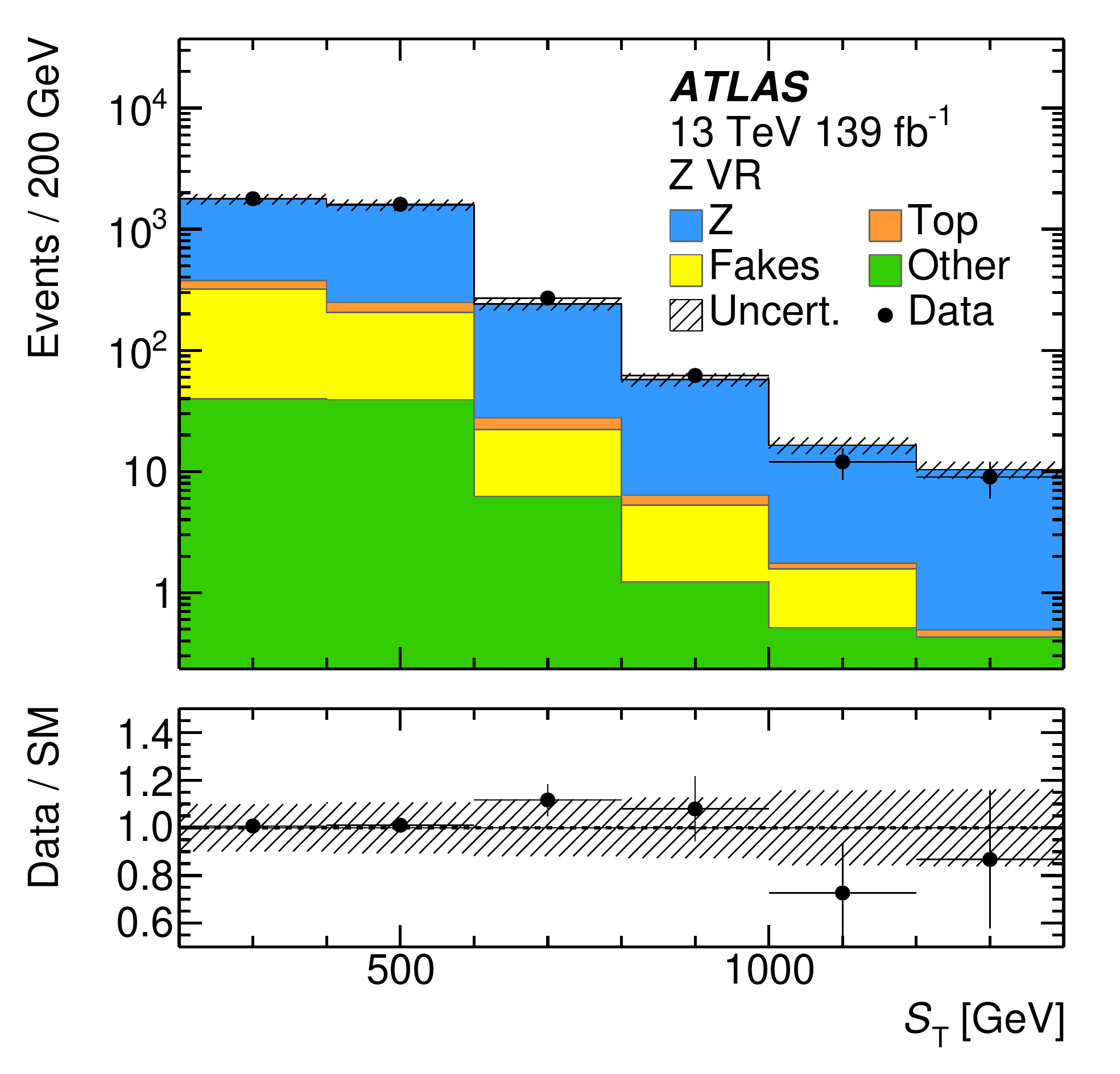}
}
\caption{Comparison of post-fit \st\ spectra with data in the \ztt\ CR and VR\@.
The hatched band corresponds to the total post-fit uncertainty, considering correlations between the individual NPs.
The ratios in the bottom panel are calculated relative to the SM prediction (background-only).
The ``Other'' template consists primarily of diboson background in the \ztt regions.
Overflow events are added to the yields in the highest \st bin.
}
\label{fig:bkgztt}
\end{figure}
 
\subsection{Top background estimate}
\label{sec:bkgttbar}
 
The top background is a sum of \ttbar and single top quark production.
It is assigned one common unconstrained normalisation factor.
In the SR, 91\% of the Top background consists of \ttbar events; the rest are single top quark events.
The single top quark contribution increases with \st to about 30\% in the last bin.
However, these fractions change by varying the NP which controls the \ttbar-\tw interference uncertainty (c.f. Section~\ref{sec:datamc}).
The admixtures of \ttbar and \tw seen in the Top CR and VR, and the SR are similar enough that any differences can be ignored.
Modelling of a possible contribution from $\ttbar+b$ events to the SR is validated in a dedicated VR, defined by the same selections as the Top CR but requiring the presence of at least three $b$-tagged jets.
Figure~\ref{fig:bkgtop} shows the post-fit \st spectra and their total uncertainties compared to data in the Top CR and VR.
 
The background-only fit described in Section~\ref{sec:bkgztt} is performed.
A good agreement between the total post-fit background prediction and data is observed.
The fitted normalisation factor is compatible with 1.
 
\begin{figure}
\centering
\subfloat[Top CR]{
\includegraphics[width=0.45\textwidth]{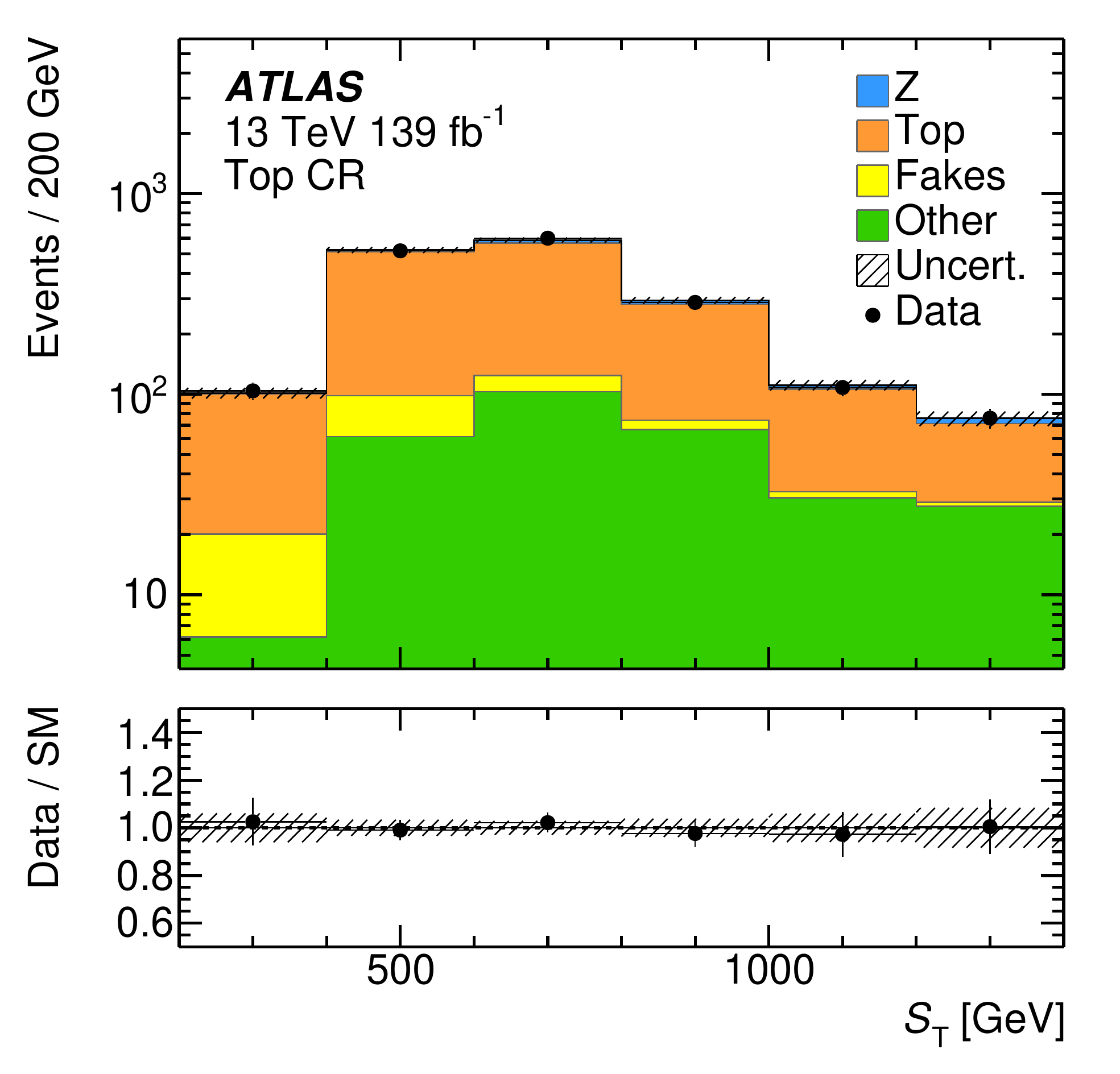}
}\quad
\subfloat[Top VR]{
\includegraphics[width=0.45\textwidth]{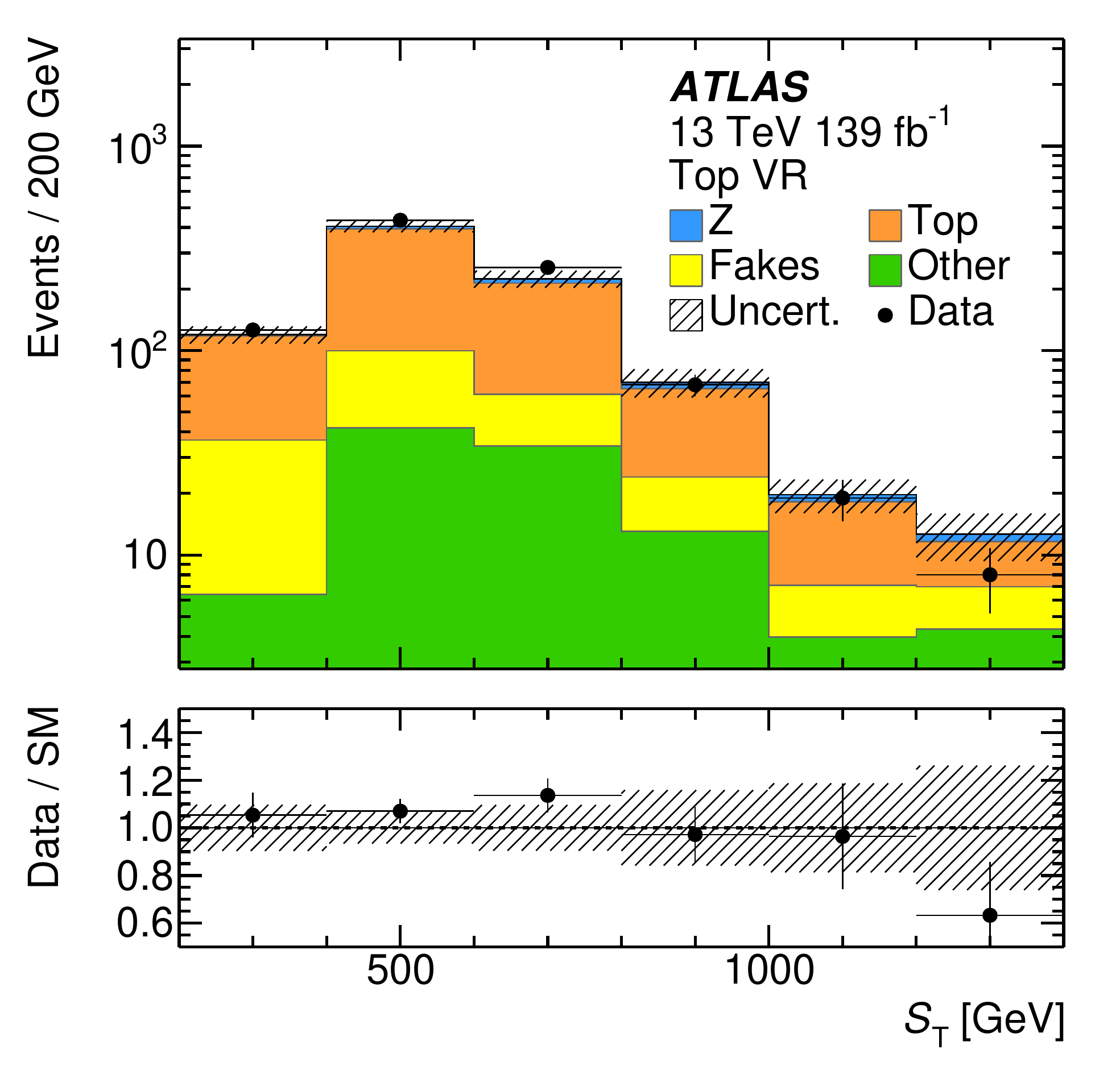}
}
\caption{Comparison of post-fit \st spectra with data in the Top CR and VR\@.
The hatched band corresponds to the total post-fit uncertainty, considering correlations between the individual NPs.
The ratios in the bottom panel are calculated relative to the SM prediction (background-only).
The ``Other'' template consists primarily of $W$ background in the Top regions.
Overflow events are added to the yields in the highest \st bin.
}
\label{fig:bkgtop}
\end{figure}
 
\subsection{Fake background estimate}
\label{sec:bkgfake}
 
This section describes the estimate of background due to jets misidentified as \tauhad.
For this purpose, a data-driven Fake Factor (FF) method is used.
 
\subsubsection{Fake background in Top CR and VR}
\label{sec:bkgfake1tau}
 
The Fake background templates in the Top CR and VR are constructed using data and MC events taken from anti-ID regions.
These regions have the same selection criteria as the corresponding analysis regions, but with the \tauhad failing its ID requirement, making the anti-ID regions enriched in Fake background.
To make events from the anti-ID regions usable as an estimate of the Fake background in the analysis regions, they must be weighted by FFs to account for different selection efficiencies between \tauhad ID and anti-ID requirements.
The FFs are measured in a region enriched in Fake background, which is orthogonal to both the analysis and anti-ID regions.
It is defined by the same selection criteria as the Top CR, but with the \etmis requirement changed to $\etmis < 30~\GeV$.
The FFs are defined as the ratio of the number of events in which \tauhad passes the ID criterion over the number of cases when it fails, expressed as a function of the \tauhad \pt, $|\eta|$ and decay mode.
Even though anti-ID and FF-measurement regions contain mostly Fake background, they also contain a small fraction of events with real \tl-leptons.
This contamination is estimated using the MC and subtracted from data event yields.
 
There are two sources of systematic uncertainty in the Fake background prediction: statistical uncertainty in the FFs and different composition of Fakes in the FF measurement, and anti-ID regions (Fakes from gluon- and quark-originated jets have different FF).
The latter uncertainty is estimated as the difference between the nominal Fake background prediction and two variations.
In the first variation, the nominal FF is replaced with an FF measured using quark-enriched Fake events from a dedicated region targeting the $\zmm+$jet process.
In the second variation, an FF measured in gluon-enriched fake events in a region designed to select mostly dijet events from pile-up collisions is used.
 
Figure~\ref{fig:fakevr1tau} shows a good agreement of the post-fit background prediction with data in the VR, defined in Section~\ref{sec:fakevr}.
In the region, 84\% of events are estimated to be Fake events.
\begin{figure}
\centering
\subfloat[Fake VR for single-\tauhad events]{
\includegraphics[width=0.45\textwidth]{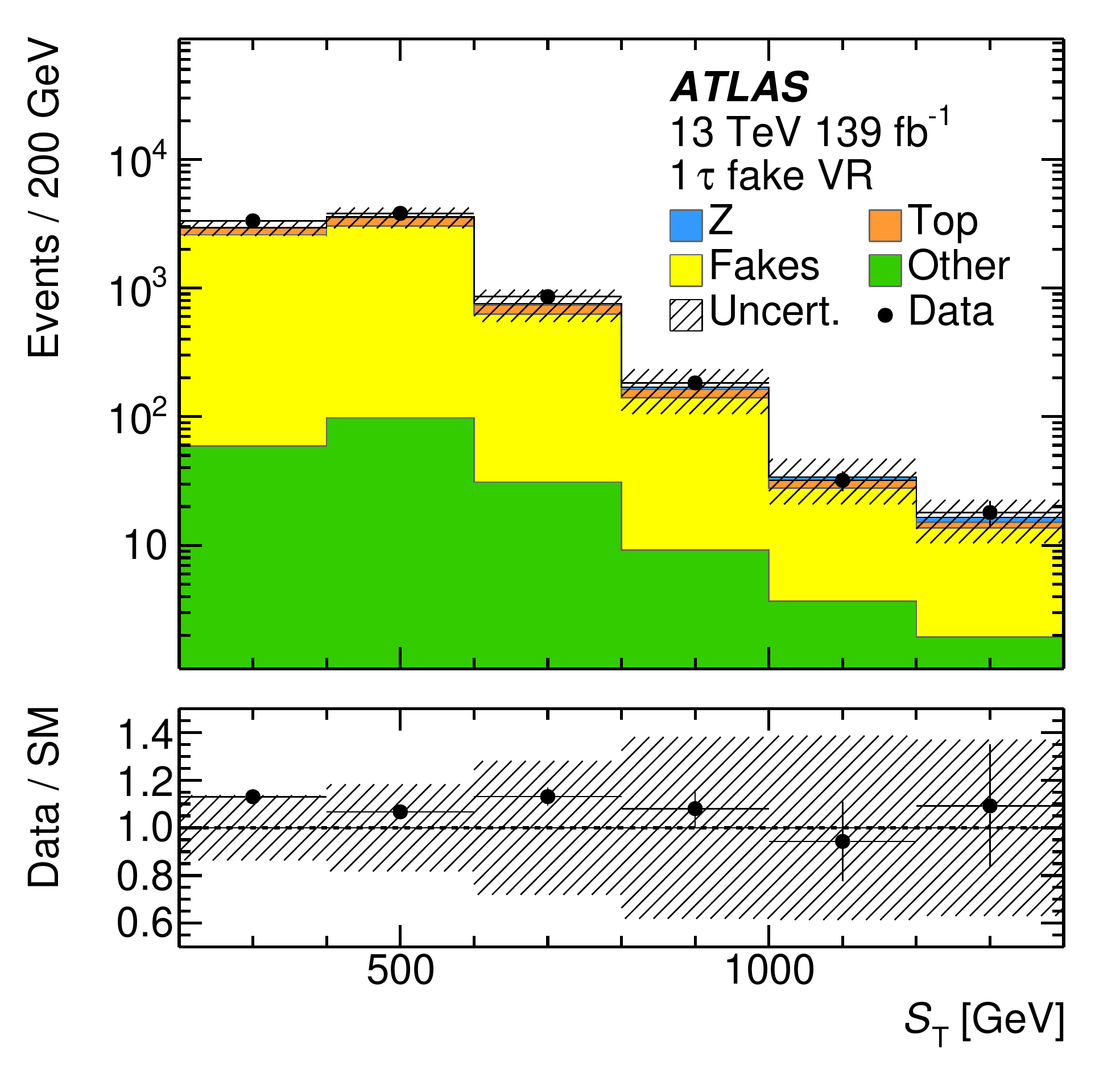}
\label{fig:fakevr1tau}
}\quad
\subfloat[Fake VR for di-\tauhad events]{
\includegraphics[width=0.45\textwidth]{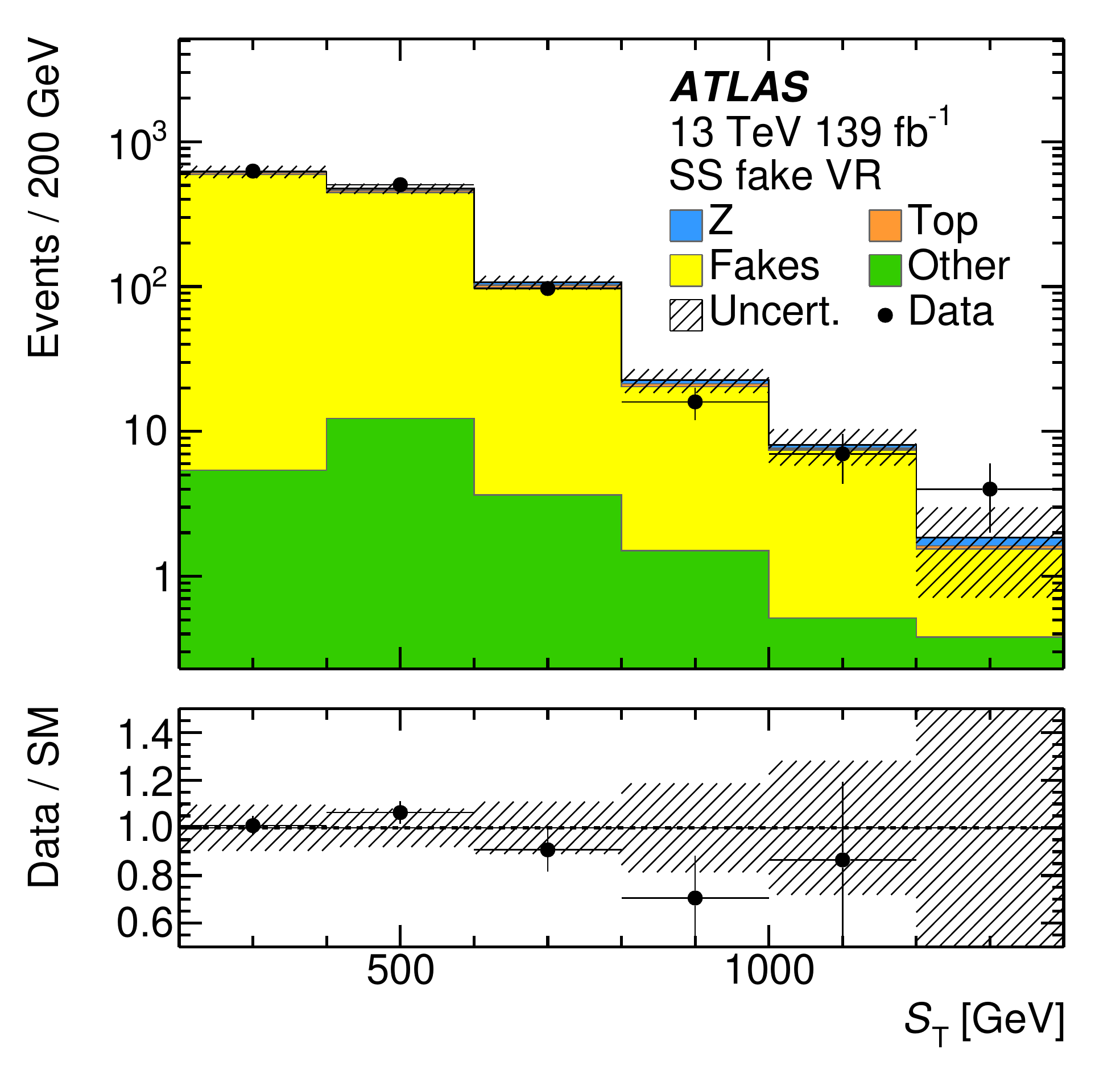}
\label{fig:fakevrhh}
}
\caption{Comparison of post-fit \st\ spectra with data in the VRs for fake background estimates.
The hatched band corresponds to the total post-fit uncertainty, considering correlations between the individual NPs.
The ratios in the bottom panel are calculated relative to the SM prediction (background-only).
The ``Other'' template consists primarily of $W$ background in the single-\tauhad fake VR and diboson background in the di-\tauhad fake VR\@.
Overflow events are added to the yields in the highest \st bin.
}
\label{fig:fakevr}
\end{figure}
 
\subsubsection{Fake background in SR and in \ztt CR and VR}
\label{sec:bkgfakehh}
 
The general idea of the fake background estimate in di-\tauhad events is similar to the one described in the previous section but with some key differences.
Because there are two $\tauhad$s in the final state, there are three types of events in the anti-ID regions, that depend on whether leading, sub-leading, or both the $\tauhad$s  fail to satisfy the \tauhad ID criterion.
The first type of events is named Leading anti-ID, the second type Sub-leading anti-ID, and the third type Double anti-ID.
Furthermore, because leading and sub-leading $\tauhad$s are selected using different \tauhad ID requirements (\textit{Medium} and \textit{Loose} ID, respectively), two sets of FFs must be measured separately for leading and sub-leading $\tauhad$s.
When the Fake background template is constructed, the FFs for \textit{Medium} ID are used for Leading anti-ID events, while the FFs for \textit{Loose} ID are used for the Sub-leading anti-ID events.
The events with both leptons failing their IDs are weighted with a product of the corresponding FFs and subtracted from the template to remove the double-counting of events with two fake $\tauhad$s.
Without this subtraction, the double-counting would occur because both the Leading and Sub-leading anti-ID events contain a fraction of events where both the $\tauhad$s are misidentified jets.
 
The FFs for di-\tauhad events are measured using events with at least one \tauhad passing the ID criterion, that further satisfy the SS charge requirement and with $\mtt<110$~\GeV or $\lt<140$~\GeV.
All the other criteria are identical to the SR, (i.e. the FF measurement region is orthogonal to the SS fake VR defined in Section~\ref{sec:fakevr}).
The FFs for the leading \tauhad are measured using events in which the subleading \tauhad passes its \tauhad ID and vice-versa for the sub-leading \tauhad FFs.
The FFs are parameterised as a function of the \tauhad \pt, $|\eta|$ and decay mode. The method for estimating the systematic uncertainties of the fake background prediction is the same as the one described in the previous section.
Figure~\ref{fig:fakevrhh} shows good agreement of the post-fit background prediction with data in the VR\@.
In the region, 93\% of events are Fake events.

% End of text imported from the .//source/bkg.tex input file

% The next lines are included from the .//source/results.tex input file
\section{Results}
\label{sec:result}
 
Figure~\ref{fig:stsr} shows data and the post-fit prediction in the SR for one \etl mass hypothesis.
The \st binning is determined by the requirement of low enough uncertainty due to the limited MC statistics.
The expected background and observed data yields are shown in Table~\ref{tab:SR}.
In ATLAS Run 2 data, 23 events are observed in the most sensitive bin ($\st > 1200$~\GeV) of the signal region, while $28 \pm 5$ background events are expected.
No excess of data over the background expectation is observed.
\begin{figure}
\centering
\includegraphics[width=0.45\textwidth]{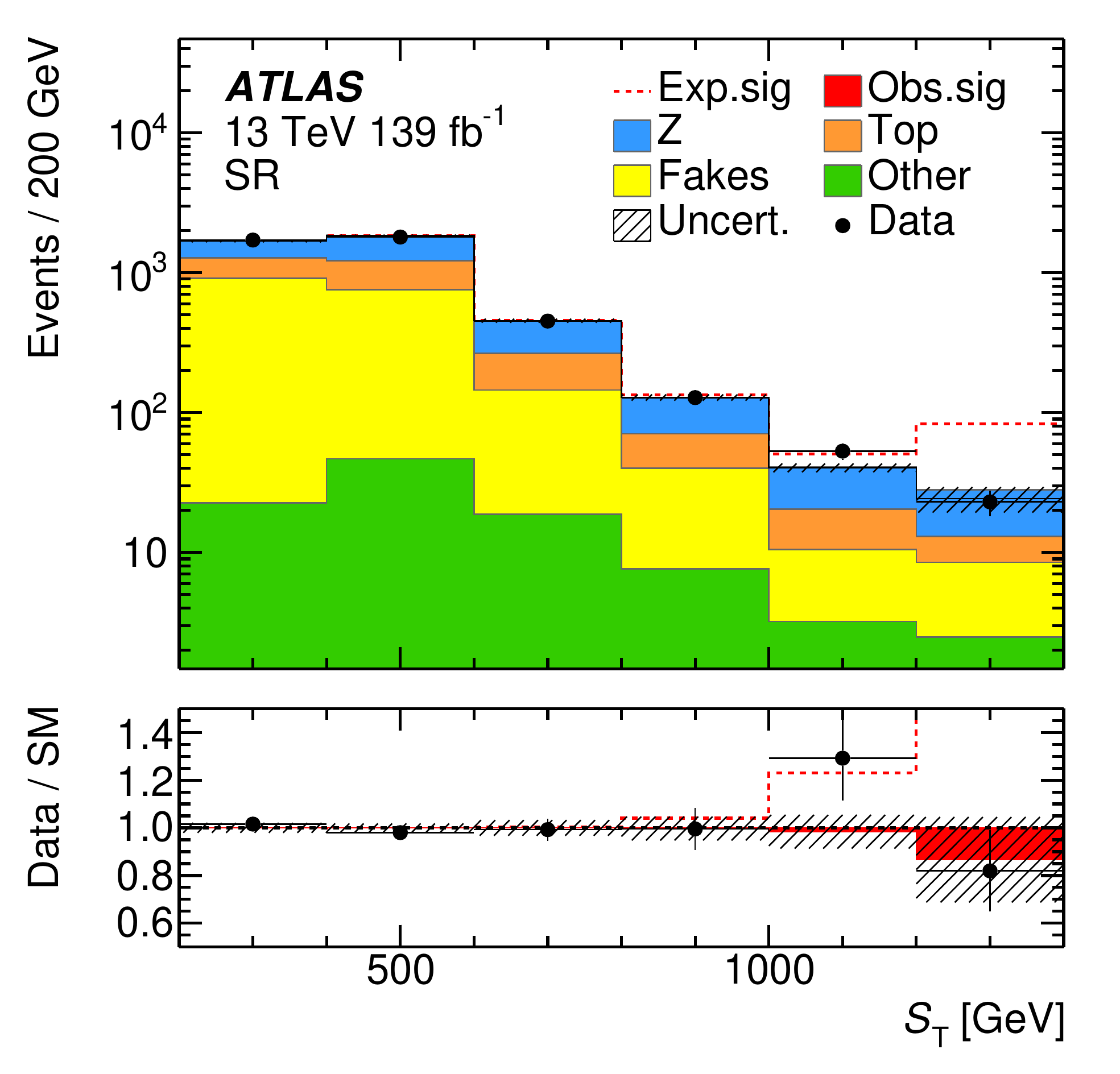}
\caption{Comparison of total -- background plus signal -- post-fit \st spectrum with data in the SR\@.
The hatched band corresponds to the total post-fit uncertainty, considering correlations between the individual NPs.
It is centered around the total post-fit prediction.
The red histogram corresponds to the signal template for \etl with a mass of 1500~\GeV and the compositeness scale set to $\Lambda=10$~\TeV.
Expected (Exp.) and best-fit (Obs.) signal templates are shown.
The ratios in the bottom panel are calculated relative to the background-only post-fit predictions (SM).
Overflow events are added to the yields in the highest \st bin.
}
\label{fig:stsr}
\end{figure}
 
\begin{table}
\centering
\caption{Post-fit yields in \st bins of the SR\@.
The fit was performed for a \etl with a mass of 1500~\GeV.
The symmetrised total post-fit uncertainty is shown in the table. }
\begin{tabular}{lccccccc}
\hline
\st range [\GeV] & Other & Fakes & Top & $Z$ & Exp. bkg. & Uncert. & Data\\
\hline
$200 - 400$     & 23  & 887  & 368  & 407  & 1685  & 35  & 1711  \\
$400 - 600$     & 47  & 709  & 464  & 615  & 1835  & 33  & 1798  \\
$600 - 800$     & 19  & 126  & 120  & 189  & 454   & 15  & 451   \\
$800 - 1000$    & 7.7 & 32.4 & 30.5 & 58.0 & 128.6 & 6.4 & 128   \\
$1000 - 1200$   & 3.2 & 7.3  & 9.9  & 20.6 & 41.0  & 2.9 & 53    \\
$>1200$         & 2.5 & 6.0  & 4.5  & 15.0 & 28.1  & 5.0 & 23    \\
\hline
\end{tabular}
\label{tab:SR}
\end{table}

Figure~\ref{fig:exctaulimitXS} shows the upper 95\%~CL limit on the \etl production cross-section as a function of \etm for a fixed value of the CI scale, $\Lambda = 10~\TeV$.
A \etl with mass less than 2.8~\TeV is excluded at 95\%~CL for this choice of $\Lambda$.
Uncertainties in the signal cross-section shown in the plot are estimated as total signal yield variations due to the MC theory uncertainties described in Section~\ref{sec:datamcsig} added in quadrature.
Figure~\ref{fig:exctaulimitLambda} shows the lower 95\%~CL limit on $\Lambda$ as a function of \etm.
The shaded area corresponds to ($\Lambda$, \etm) points where the interaction certainly cannot be treated as an effective four-fermion contact interaction~\cite{PhysRevD.42.815}.
Its boundary is given by a line $\Lambda = \etm$.
The intersection of this line with the observed limit gives an upper 95\%~CL limit on \etm of 4.6~\TeV for a scenario with $\Lambda = \etm$.
\begin{figure}
\centering
\subfloat[Cross-section limit, $\Lambda = 10~\TeV$.]{
\includegraphics[width=0.45\textwidth]{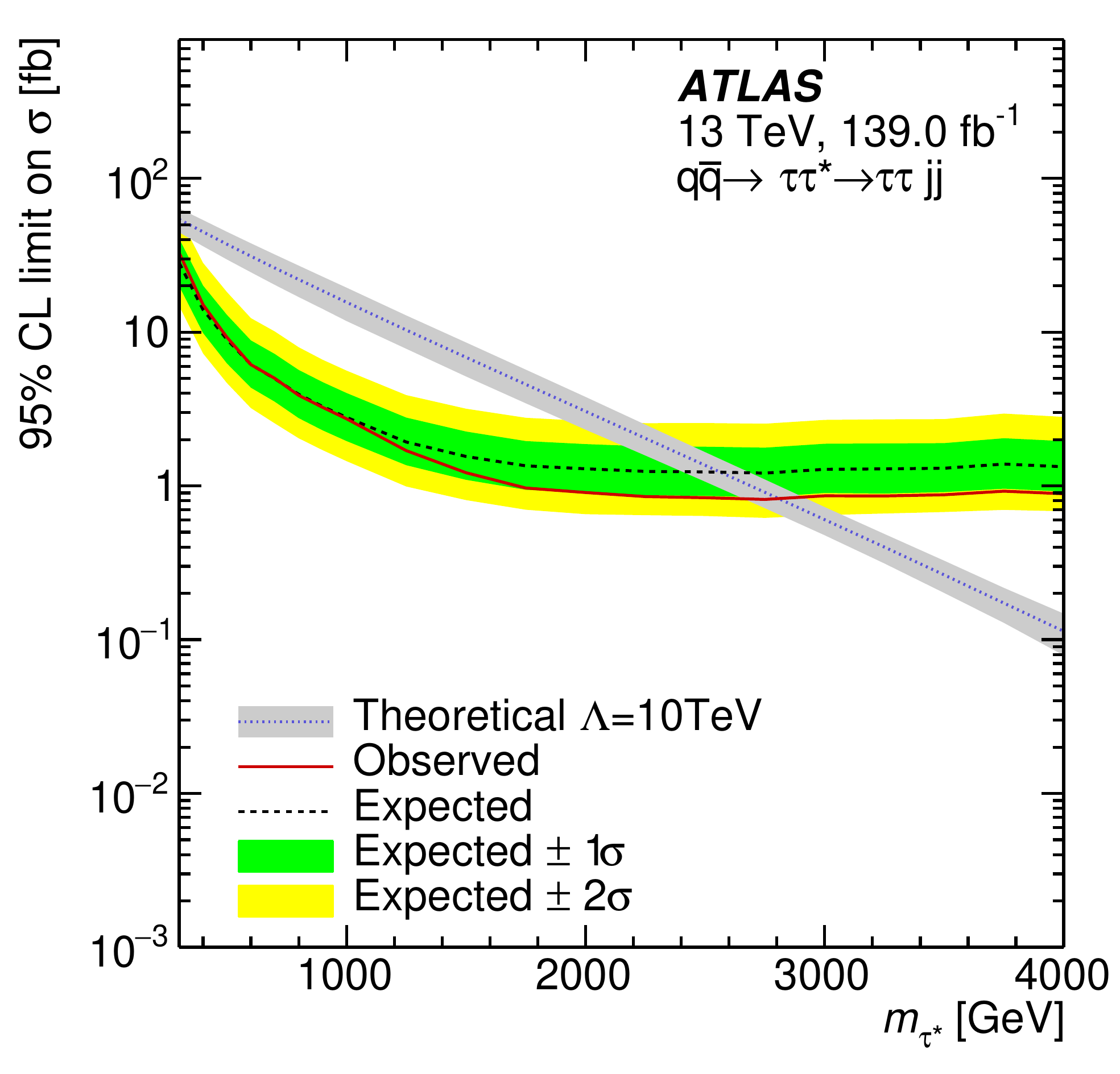}
\label{fig:exctaulimitXS}
}\quad
\subfloat[$\Lambda$ limit, $\Lambda = \etm$.]{
\includegraphics[width=0.45\textwidth]{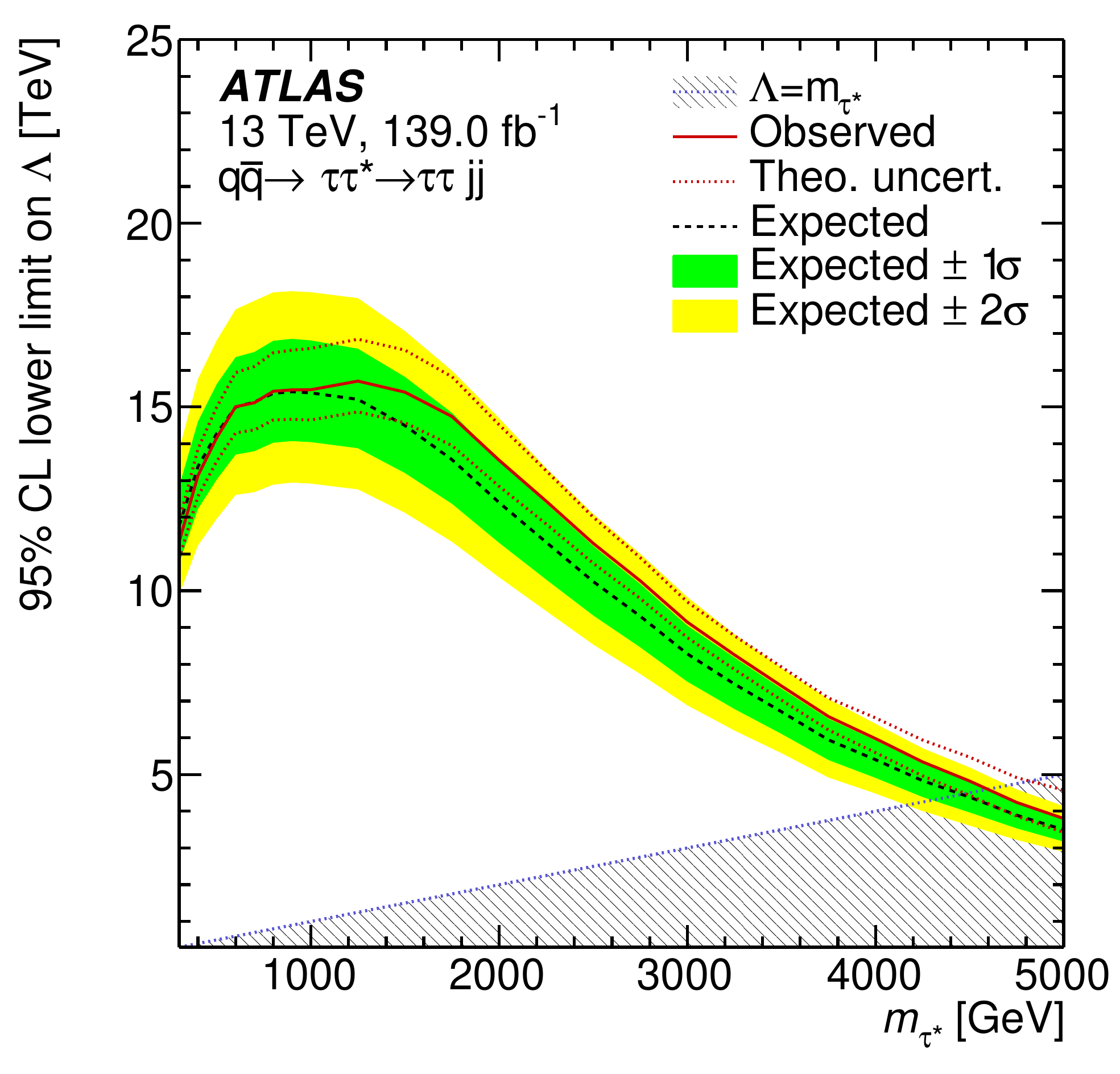}
\label{fig:exctaulimitLambda}
}
\caption{95\%~CL limits on selected \etl model hypotheses.
Figure~\ref{fig:exctaulimitXS} shows the upper 95\%~CL limit on the \etl production cross-section as a function of \etm for a fixed value of the CI scale, $\Lambda = 10~\TeV$.
Figure~\ref{fig:exctaulimitLambda} shows the lower 95\%~CL limit on $\Lambda$ as a function of \etm.
The shaded area corresponds to ($\Lambda$, \etm) points where the interaction cannot be treated as an effective four-fermion contact interaction~\cite{PhysRevD.42.815}.
The observed and expected limits are shown by the solid red and dashed black lines, respectively.
Boundaries of the green (yellow) band display $\pm 1\sigma$ ($\pm 2\sigma$) statistical uncertainty in the expected limit.
The dotted blue line and the grey band in Figure~\ref{fig:exctaulimitXS} display the \etl production cross-section and its uncertainty.
The dotted red lines in Figure~\ref{fig:exctaulimitLambda} show observed limits on $\Lambda$ given $\pm 1\sigma$ variations of the signal production cross-section.
}
\label{fig:exctaulimit}
\end{figure}
 
Figure~\ref{fig:lqlimit} shows the upper 95\%~CL limit on the LQ production cross-section as a function of mass.
Signal cross-sections displayed in the limit plot are computed at approximate NNLO, as described in Sec.~\ref{sec:datamcsig}.
Leptoquarks with masses below 1.3~\TeV are excluded at 95\% CL.
\begin{figure}
\centering
\includegraphics[width=0.45\textwidth]{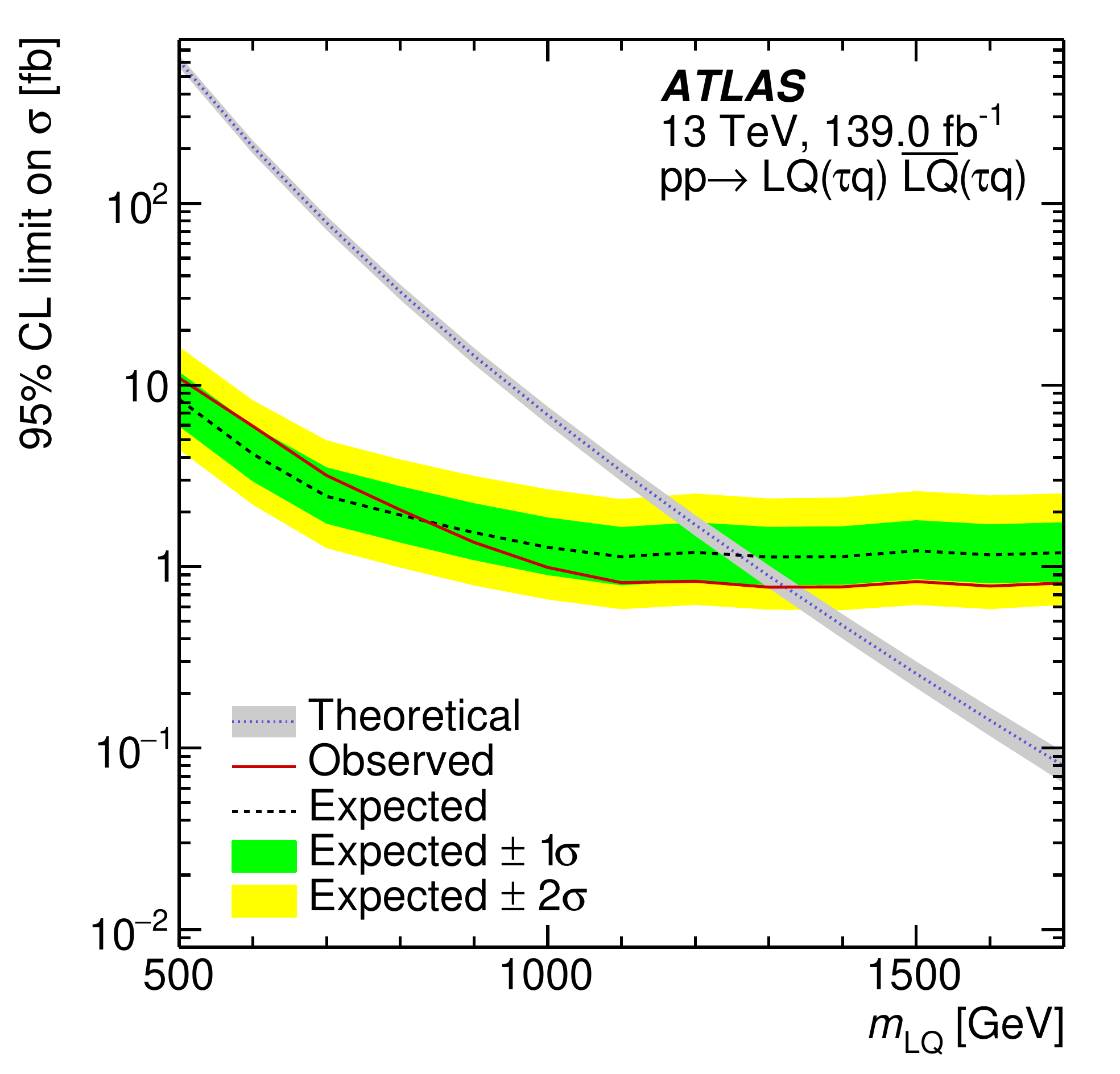}
\caption{Upper 95\%~CL limit on the LQ pair production cross-section as a function of mass.
The LQ decays into a \tl-lepton and a $c$-quark (or a lighter flavour quark) with a BR of 1.
The observed and expected limits are shown by the solid red and dashed black lines, respectively.
Boundaries of the green (yellow) band display $\pm 1\sigma$ ($\pm 2\sigma$) statistical uncertainty in the expected limit.
The dotted blue line and the grey band display the LQ production cross-section and its uncertainty.
}
\label{fig:lqlimit}
\end{figure}

% End of text imported from the .//source/results.tex input file

\FloatBarrier

% The next lines are included from the .//source/conclusion.tex input file
\section{Conclusion}
\label{sec:conclusion}
 
A search for \etl or LQ in events with two $\tauhad$s and two or more jets was performed using \textit{pp}~collision data at $\sqrt{s} = 13~\tev$ recorded by the ATLAS experiment during the LHC Run 2 in 2015--2018.
The total integrated luminosity is 139~\ifb.
The \etl is assumed to be produced and decayed via a four-fermion CI with a \tl-lepton and a \qqbar pair.
The hypothetical LQ is assumed to couple to a $c$-quark and a \tl-lepton and is produced together with its antiparticle via the strong interaction.
The BR of the LQ decay is 1.
The main backgrounds to the analysis are \ztt and Top production processes, and Fake events.
The background is estimated with MC, except for the Fake events whose yields are predicted with a data-driven FF method.
No excess of data over the background prediction is observed.
\etl with masses below 2.8~\tev and 4.6~\tev are excluded at 95\% CL in scenarios with the CI scale $\Lambda$ set to 10~\tev and to \etm, respectively.
For comparison, the previous ATLAS mass limit is 2.5~\tev in the regime $\etm = \Lambda$~\cite{EXOT-2012-20}.
LQs with masses below 1.3~\tev are excluded at 95\% CL, assuming the branching ratio for their decay into the $c$-quark--\tl-lepton pair is equal to one.
It is the first ATLAS search for such a LQ decay scenario.
The analysis does not exploit jet flavour-tagging in the signal region, and the limits hold for hypotheses of a LQ coupling to any lighter quark flavour and the \tl-lepton.

% End of text imported from the .//source/conclusion.tex input file

\section*{Acknowledgements}

% The next lines are included from the .//acknowledgements/Acknowledgements.tex input file

We thank CERN for the very successful operation of the LHC, as well as the
support staff from our institutions without whom ATLAS could not be
operated efficiently.
 
We acknowledge the support of
ANPCyT, Argentina;
YerPhI, Armenia;
ARC, Australia;
BMWFW and FWF, Austria;
ANAS, Azerbaijan;
CNPq and FAPESP, Brazil;
NSERC, NRC and CFI, Canada;
CERN;
ANID, Chile;
CAS, MOST and NSFC, China;
Minciencias, Colombia;
MEYS CR, Czech Republic;
DNRF and DNSRC, Denmark;
IN2P3-CNRS and CEA-DRF/IRFU, France;
SRNSFG, Georgia;
BMBF, HGF and MPG, Germany;
GSRI, Greece;
RGC and Hong Kong SAR, China;
ISF and Benoziyo Center, Israel;
INFN, Italy;
MEXT and JSPS, Japan;
CNRST, Morocco;
NWO, Netherlands;
RCN, Norway;
MEiN, Poland;
FCT, Portugal;
MNE/IFA, Romania;
MESTD, Serbia;
MSSR, Slovakia;
ARRS and MIZ\v{S}, Slovenia;
DSI/NRF, South Africa;
MICINN, Spain;
SRC and Wallenberg Foundation, Sweden;
SERI, SNSF and Cantons of Bern and Geneva, Switzerland;
MOST, Taiwan;
TENMAK, T\"urkiye;
STFC, United Kingdom;
DOE and NSF, United States of America.
In addition, individual groups and members have received support from
BCKDF, CANARIE, Compute Canada and CRC, Canada;
PRIMUS 21/SCI/017 and UNCE SCI/013, Czech Republic;
COST, ERC, ERDF, Horizon 2020 and Marie Sk{\l}odowska-Curie Actions, European Union;
Investissements d'Avenir Labex, Investissements d'Avenir Idex and ANR, France;
DFG and AvH Foundation, Germany;
Herakleitos, Thales and Aristeia programmes co-financed by EU-ESF and the Greek NSRF, Greece;
BSF-NSF and MINERVA, Israel;
Norwegian Financial Mechanism 2014-2021, Norway;
NCN and NAWA, Poland;
La Caixa Banking Foundation, CERCA Programme Generalitat de Catalunya and PROMETEO and GenT Programmes Generalitat Valenciana, Spain;
G\"{o}ran Gustafssons Stiftelse, Sweden;
The Royal Society and Leverhulme Trust, United Kingdom.
 
The crucial computing support from all WLCG partners is acknowledged gratefully, in particular from CERN, the ATLAS Tier-1 facilities at TRIUMF (Canada), NDGF (Denmark, Norway, Sweden), CC-IN2P3 (France), KIT/GridKA (Germany), INFN-CNAF (Italy), NL-T1 (Netherlands), PIC (Spain), ASGC (Taiwan), RAL (UK) and BNL (USA), the Tier-2 facilities worldwide and large non-WLCG resource providers. Major contributors of computing resources are listed in Ref.~\cite{ATL-SOFT-PUB-2021-003}.

% End of text imported from the .//acknowledgements/Acknowledgements.tex input file

\printbibliography
 
\clearpage
% ATLAS Collaboration author list
% Reference date of EXOT-2020-18 is 2022-12-16
% Author list last updated on date 28-JUN-23
% Data extracted on 28-Jun-2023 for paper reference EXOT-2020-18
% at 3:36pm
 
\begin{flushleft}
\hypersetup{urlcolor=black}
{\Large The ATLAS Collaboration}

\bigskip

\AtlasOrcid[0000-0002-6665-4934]{G.~Aad}$^\textrm{\scriptsize 102}$,
\AtlasOrcid[0000-0002-5888-2734]{B.~Abbott}$^\textrm{\scriptsize 120}$,
\AtlasOrcid[0000-0002-1002-1652]{K.~Abeling}$^\textrm{\scriptsize 55}$,
\AtlasOrcid[0000-0001-5763-2760]{N.J.~Abicht}$^\textrm{\scriptsize 49}$,
\AtlasOrcid[0000-0002-8496-9294]{S.H.~Abidi}$^\textrm{\scriptsize 29}$,
\AtlasOrcid[0000-0002-9987-2292]{A.~Aboulhorma}$^\textrm{\scriptsize 35e}$,
\AtlasOrcid[0000-0001-5329-6640]{H.~Abramowicz}$^\textrm{\scriptsize 151}$,
\AtlasOrcid[0000-0002-1599-2896]{H.~Abreu}$^\textrm{\scriptsize 150}$,
\AtlasOrcid[0000-0003-0403-3697]{Y.~Abulaiti}$^\textrm{\scriptsize 117}$,
\AtlasOrcid[0000-0003-0762-7204]{A.C.~Abusleme~Hoffman}$^\textrm{\scriptsize 137a}$
% ...
\bigskip
\\

% ...
$^{29}$Physics Department, Brookhaven National Laboratory, Upton NY; United States of America.\\
% ...
$^{35}$$^{(a)}$Facult\'e des Sciences Ain Chock, R\'eseau Universitaire de Physique des Hautes Energies - Universit\'e Hassan II, Casablanca;$^{(b)}$Facult\'{e} des Sciences, Universit\'{e} Ibn-Tofail, K\'{e}nitra;$^{(c)}$Facult\'e des Sciences Semlalia, Universit\'e Cadi Ayyad, LPHEA-Marrakech;$^{(d)}$LPMR, Facult\'e des Sciences, Universit\'e Mohamed Premier, Oujda;$^{(e)}$Facult\'e des sciences, Universit\'e Mohammed V, Rabat;$^{(f)}$Institute of Applied Physics, Mohammed VI Polytechnic University, Ben Guerir; Morocco.\\
% ...
$^{49}$Fakult\"{a}t Physik , Technische Universit{\"a}t Dortmund, Dortmund; Germany.\\
% ...
$^{55}$II. Physikalisches Institut, Georg-August-Universit\"{a}t G\"ottingen, G\"ottingen; Germany.\\
% ...
$^{102}$CPPM, Aix-Marseille Universit\'e, CNRS/IN2P3, Marseille; France.\\
% ...
$^{117}$Department of Physics, New York University, New York NY; United States of America.\\
% ...
$^{120}$Homer L. Dodge Department of Physics and Astronomy, University of Oklahoma, Norman OK; United States of America.\\
% ...
$^{137}$$^{(a)}$Departamento de F\'isica, Pontificia Universidad Cat\'olica de Chile, Santiago;$^{(b)}$Millennium Institute for Subatomic physics at high energy frontier (SAPHIR), Santiago;$^{(c)}$Instituto de Investigaci\'on Multidisciplinario en Ciencia y Tecnolog\'ia, y Departamento de F\'isica, Universidad de La Serena;$^{(d)}$Universidad Andres Bello, Department of Physics, Santiago;$^{(e)}$Instituto de Alta Investigaci\'on, Universidad de Tarapac\'a, Arica;$^{(f)}$Departamento de F\'isica, Universidad T\'ecnica Federico Santa Mar\'ia, Valpara\'iso; Chile.\\
% ...
$^{150}$Department of Physics, Technion, Israel Institute of Technology, Haifa; Israel.\\
$^{151}$Raymond and Beverly Sackler School of Physics and Astronomy, Tel Aviv University, Tel Aviv; Israel.\\
% ...
$^{a}$ Also Affiliated with an institute covered by a cooperation agreement with CERN.\\
$^{b}$ Also at An-Najah National University, Nablus; Palestine.\\
$^{c}$ Also at Borough of Manhattan Community College, City University of New York, New York NY; United States of America.\\
$^{d}$ Also at Bruno Kessler Foundation, Trento; Italy.\\
$^{e}$ Also at Center for High Energy Physics, Peking University; China.\\
$^{f}$ Also at Center for Interdisciplinary Research and Innovation (CIRI-AUTH), Thessaloniki ; Greece.\\
% ...

\end{flushleft}

% Created with Glance <Atlas.Glance@cern.ch>

\end{document}